\documentstyle[epsf,psfig,rotate,amstex,amssymb]{l-aa} % LaTeX A&A  Monotype Times Fonts

\begin{document}

   \thesaurus{08         % A&A Section 8: Stars
              (08.02.4;  % Stars: binaries: spectroscopic
               08.03.2;  % Stars: chemically peculiar
               08.05.3;  % Stars: evolution
               08.18.1;  % Stars: rotation
               08.18.1)  % Stars: $\delta$ Scuti
             }

\title{Are metallic A-F giants evolved Am stars?\\
Rotation and rate of binaries among giant F stars \thanks{Based on observations 
collected at Observatoire de Haute Provence (OHP), France.}$^,$
\thanks{Tables 2, 3, 6 are also available in electronic form at CDS via
anonymous ftp to cdsarc.u-strasbg.fr (130.79.128.5) or via
http://cdsweb.u-strasbg.fr/Abstract.html}}

\author{M. K\"unzli \and P. North}

\offprints{P. North}

\institute{Institut d'Astronomie de l'Universit\'e de Lausanne,
 CH-1290 Chavannes-des-Bois, Switzerland
 }
\date{Received 3.12.96; accepted 25.5.97 }

\maketitle

\markboth{M. K\"unzli et al.: Are metallic giants evolved Am stars?}
{M. K\"unzli et al.: Are metallic giants evolved Am stars?}

\begin{abstract}

We test the hypothesis of Berthet (1992) which foresees that Am stars become
giant metallic A and F stars (defined by an enhanced value of the blanketing
parameter $\Delta m_2$ of the Geneva photometry) when they evolve.  

If this hypothesis is right, Am and metallic A-FIII stars need to have the
 same rate of binaries and a similar distribution of $v\sin i$. From our new
spectroscopic data and from $v\sin i$ and radial velocities in the literature,
we show that it is not the case. The metallic giant stars are often  fast 
rotators with $v\sin i$ larger than 100 km\,s$^{-1}$, while the maximum
rotational velocity for Am stars is about 100 km\,s$^{-1}$. The rate of tight
binaries  with periods less than 1000 days is less than 30 \% among metallic
giants, which is incompatible with the value of 75 \% for Am stars
(Abt \& Levy 1985). Therefore, the simplest way to explain the existence of
giant metallic F stars is to suggest that all normal A and early F stars might
go through a short "metallic" phase when they are finishing their life on the
main sequence.

Besides, it is shown that only giant stars with spectral type comprised
between F0 and F6 may have a really enhanced $\Delta m_2$ value, while all
A-type giants seem to be normal. 

\end{abstract}

\section{Introduction}
Main sequence A and F stars show various interesting phenomena: chemical
peculiarities (Am,Fm,Ap...)  on the one hand and/or pulsation  on
the other hand. Numerous studies have been devoted to this part of the HR
diagram near the ZAMS, but few have explored more evolved stars. In this
paper, we examine the rate of binaries and rotation of giant A and F stars.

In a study of 132 giant A and F stars, Hauck (1986) showed that 36 \% of
them have an enhanced value of the blanketing parameter $\Delta m_2$
(defined roughly speaking by $m_2(observed)-m_2(Hyades)$) characterising Am and
$\delta$ Del stars. He showed also that this parameter could probably be
interpreted in terms of metallicity for giant A and F stars. This interpretation
was confirmed by Berthet (1990, 
1991) from a detailed abundance analysis of such objects. His study
pointed out that these stars have chemical properties similar to those
of $\delta$ Del stars, i.e. an overabundance of iron peak elements and
especially of heavier elements such as Sr and Ba, and solar composition
for Ca and Sc. Am stars have similar characteristics for iron peak
 elements, but Ca and Sc are deficient by factors of about 10.

Based on chemical abundances, position in the $(\beta,M_v)$ plane of
Str\"omgren photometry, rotation and duplicity
 of $\delta$ Del stars,
Kurtz (1976) suggested that these stars are evolved Am stars. The
fact that $\delta$ Del and giant metallic A and F stars harbour
similar chemical properties suggests also a link between Am and
giant metallic A and F stars. Thus, a star could begin its life
on the main sequence as an Am star, then, as abundances of Ca
and Sc increase to quasi-solar values with evolution, it would become
a $\delta$ Del star and finally a giant metallic A-F where Ca and
Sc are solar (Berthet 1992).

To explain the emergence of chemical anomalies in Am stars, one generally
invokes the radiative diffusion theory developed by Michaud et al (1983).
This theory predicts that, in slow rotators, helium is no longer
sustained and flows inside the star and gradually disappears from the atmosphere.
The diffusion process could therefore take place just below the thin H
convective zone where the diffusion time is short with respect
to the stellar lifetime; the chemical elements whose
radiative acceleration is larger than gravity become overabundant
and, in the opposite case, underabundant. The convective zone becomes
deeper with evolution and so leads to normalisation of the surface
abundance. This theory explains the chemical anomalies in Am stars
and the increase of Ca and Sc abundances with time.

The chemical anomalies predicted by the diffusion theory are generally
larger than the observed anomalies. These differences come from uncertainties of
the atomic data but also from some mechanism in the stellar atmosphere. The mass
loss probably takes a prominent part: a rate of about $10^{-15}M_\odot /year$ is
sufficient to reduce the theoretical overabundance of heavy elements to observed
abundances (Charbonneau \& Michaud 1991). Fast rotation
($v\sin i$ $>$ 120 km\,s$^{-1}$) generates meridional circulation which prevents
the disappearance of the helium ionisation zone, so the metallic anomalies
cannot appear (Michaud 1983). Observationally, the upper limit of rotation for
Am stars is about 100 km\,s$^{-1}$ (Abt \& Moyd 1973), while for normal stars,
we observe projected rotational velocities larger than 100 km\,s$^{-1}$, which
is in agreement with the diffusion theory. The meridional circulation cannot be 
invoked to reduce theoretical abundances to observed ones: in some cases, they
can increase the theoretical abundances and  qualitatively no relation exists
between $v\sin i$ and the chemical anomalies in the velocity range
characteristic of Am and Fm stars (0-100 km\,s$^{-1}$). Indeed, the time
diffusion under the hydrogen convective zone is shorter than that of the
meridional circulation.

The rate of binaries is completely different in Am and normal stars: Am stars
are often members of tight binaries ($P \leq$ 100 days), while normal stars in
double systems often have larger periods ($P >$ 100 days). So, we would be
tempted to explain the low rotation of Am stars by tidal braking. Nevertheless,
Zahn (1977) showed that this effect is really important only for tight binaries
with a period less than 7 days. On the other hand, Abt \& Levy (1985) showed
that 75 \% of Am stars have periods below 1000 days, so they do not necessarily
belong to tight binaries. Other mechanisms must contribute to reduce the
rotational velocity to 100 km\,s$^{-1}$ or less. One of them may be the
evolutionary expansion of stars during their main sequence lifetime. According
to Abt \& Levy (1985), during this phase, the velocity decreases by a factor of
2 and consequently most of the normal A stars may become Am stars before leaving
the main sequence. These authors suggest also the possibility of tidal braking
during the pre-main sequence phase to explain the exclusion of normal stars with
a period comprised between 10 and 100 days.

The goal of the present work is to compare the $v\sin i$ and the rate of
binaries among Am and giant metallic A and F stars in order to consider the
possibility of a link between these two types of stars. To this end, we have
measured at OHP some of Hauck's stars having no radial velocities or $v\sin i$,
or only old determinations. These data should allow to strengthen a preliminary
work (North 1994) which casts some doubts on the validity of the scenario
advocated by Berthet (1992).

\section{The sample and the observations}

We defined the sample to be observed from Table 3 and Table 4 of Hauck (1986).
The reason why we considered Table 4 (containing stars classified
spectroscopically as dwarfs but photometrically as giants) was that the
photometric criterion of luminosity seems much more reliable than the
spectroscopic luminosity class.

The selection criteria were:
\begin{itemize}
\item[-]Visibility from OHP ($\delta \gtrsim -20^o$)
\item[-]Insufficient $V_r$ data or unknown $v\sin i$
\end{itemize}

By insufficient $V_r$, we mean that the star has less than three published
$V_r$ values in the literature. Applying these criteria, 50 A and F giant stars
were selected, both normal and metallic. It was judged useful to have a good
estimate of the binary freqency of normal A and F giants for reference purposes.
40 stars in the sample are non-metallic, while only ten are metallic according
to the criterion $\Delta m_2 \geq 0.013$ (see section 5.1). Thus, the estimate
of binary frequency among metallic giants will mainly rely upon old, published
$V_r$ data. The $\Delta m_2$ values of the sample stars range between  -0.027
and 0.082. They have been updated using a new reference sequence in the
$m_2/B2-V1$ diagram (Hauck et al. 1991), and sometimes by complementary
photometric measurements. Eight stars belong to the $\delta$ Scuti class, and
nine are spectroscopically classified as dwarf but have a large $\Delta d$
parameter ($\Delta d$ is the luminosity parameter of Geneva photometry,
equivalent to Str\"omgren's $\delta c_1$) indicating that they very probably are
real giants. All stars are bright, with $V \leq 7.0$, implying that interstellar
reddening is insignificant.

The observations were performed at the Observatoire de Haute Provence (OHP) with
the Aurélie spectrograph attached to the 1.52m telescope at the Coudé focus
(Gillet et al 1994) in 1994. Four runs of a few nights' duration each were made
respectively at the beginning of May and June for 18 stars and at the beginning
of November and December for 32 stars. These measurements should allow to
detect binary stars with a small period ($P \lesssim$ 100 days). 

The detector is a double barrette CCD Thomson TH7832 with 2048 pixels having a
size of 750 x 13 $\mu$m. The spectra were obtained at a reciprocal dispersion
of 8 \AA $mm^{-1}$ in the spectral region centred on H$\beta$
[4780 \AA, 5000 \AA]. The reduction was made at the OHP with IHAP procedures,
using comparison spectra of thorium. Each stellar measurement was preceded by
a calibration exposure to compensate for instrumental drift as much as possible.
To normalise our spectrum, we simply fit a straight line to the continuum.
291 stellar exposures were made, of which 54 were devoted to standard stars.
For most of the spectra a signal-to-noise of 150 was achieved. The minimum
number of exposures per star was 2 and the maximum 11, with a mean value of
5 or 6 measurements per star.

\section{Radial velocity determinations}
\subsection{$H\beta$ fitted by a lorentz profile.}

Some stars of the sample rotate very fast and some are very hot, with effective
temperature of about 8000 K. Therefore, these stars cannot be measured with an
instrument such as Coravel, which needs many narrow lines. The adopted technique
consists in fitting a lorentz profile on  the $H\beta$ line [4861.331 \AA]
minimising the $\chi^2$, i.e. the differences between the lorentz profile and
the observed line. The advantage of a lorentz function over a gaussian is to be
more peaked at the centre, so hydrogen lines are better fitted.

A lorentz profile is defined by 3 parameters (Equation 1): $\lambda_0$ is the
centre, $i_0$ the intensity or height and $b$ the half-intensity width of the
profile. 
\begin{equation}
L(\lambda)=\frac{i_0b^2}{(\lambda-\lambda_0)^2+b^2}
\end{equation}
The three parameters of $L(\lambda)$ are fitted by least-squares; meanwhile the
optimum choice will depend on the limits fixing the spectral range on which the
fit is made. To avoid too subjective a choice, we fix on both sides of the line
two points determining a segment on which the limit is randomly chosen
(Figure 1). Thus, we generate 100 lorentz profiles with different limits on
each $H\beta$ line. A profile is taken into account only if the $\chi^2$ value
does not exceed 0.05 in order to avoid hazardous fits. Then, we define for each
line $H\beta$ the mean value of the central wavelength and the dispersion:
\begin{equation}
\overline{\lambda_0(H\beta)}=\frac{1}{n}\sum_{i=i}^{n}\lambda_0(H\beta)_i ,
n = 100
\end{equation}
\begin{equation}
\sigma_{\lambda_0(H\beta)}=\sqrt{\sum_{i=i}^{n}\frac{1}{n}
(\lambda_0(H\beta)_i-\overline{\lambda_0(H\beta)})^2} , n = 100
\end{equation}

The dispersion $\sigma_{\lambda_0(H\beta)}$ depends on the physical properties
of the star, essentially the temperature, the surface gravity and the rotation
but also the quality of the spectrum, i.e. the signal-to-noise. We have chosen
to distribute the $V_{r}$ measurements into categories of decreasing discrete
precision on the basis of $\sigma_{\lambda_0(H\beta)}$ values (see Table 1):
measurements with $\sigma_{\lambda_0(H\beta)}$ between 0 and 3 m\AA\ will be
considered as having $\sigma_{V_{r}}^s  = 0.19$ km\,s$^{-1}$, those with
$\sigma_{\lambda_0(H\beta)}$ between 3 and 6 m\AA\ will have
$\sigma_{V_{r}}^s = 0.38 $ km\,s$^{-1}$, etc. The value of 3 m\AA\ is completely
 arbitrary, corresponding to a radial velocity of about 0.2 km\,s$^{-1}$.
The $\sigma_{\lambda_0(H\beta)}$ and the central wavelength
$\overline{\lambda_0(H\beta)}$ also depend on the choice of the limits for the
fit. This effect is discussed in detail in appendix A.

\begin{table}
\hspace{0cm}{\bf Table 1.} Dispersion in radial velocity $\sigma_{V_{r}}^s$
from  the dispersion in wavelength $\sigma_{\lambda_0(H\beta)}$.
\begin{center}
\begin{tabular}{l c}
\hline\multicolumn{1}{l}{$\sigma_{\lambda_0(H\beta)}$ [\AA]}
     &\multicolumn{1}{c}{$\sigma_{V_{r}}^s$ [km\,s$^{-1}$]} \\
\hline
\ [0.000, 0.003] & 0.19 \\ 
\ [0.003, 0.006] & 0.38 \\
\ [0.006, 0.009] & 0.56 \\
\ [0.009, 0.012] & 0.74 \\
...\\
\hline
\end{tabular}
\end{center}
\end{table}

The term $\sigma_{V_{r}}^s$ is defined as the internal error due to the
effective temperature, the surface gravity, the rotation and the $S/N$ ratio.
The total internal error includes an additional term due to the instrumental
drift during the night. This variability was similar for the four missions and
Figures 2 and 3 show the fluctuations of the position of the thorium lines
during the night of 6 and 7 November and during the night of 5 and 6 December
1994 for 3 lines. When we filled up the nitrogen tank, at the beginning and in
the middle of the night, a strong variation  appeared as shown on these figures.
Between these jumps, we observe a regular variation of about 0.12 pixels per
hour. We therefore have a variation of 0.16 km\,s$^{-1}$ per 12 minutes which
corresponds to the average time of stellar exposure. For this reason, the
internal error due to this factor ($\sigma_{V_{r}}^t$) is estimated at
0.16 km\,s$^{-1}$. Finally, the total internal error is written by
$I=\sqrt{(\sigma_{V_{r}}^s)^2+(\sigma_{V_{r}}^t)^2}$.

\begin{figure}
\epsfysize=9cm
\leavevmode\epsffile{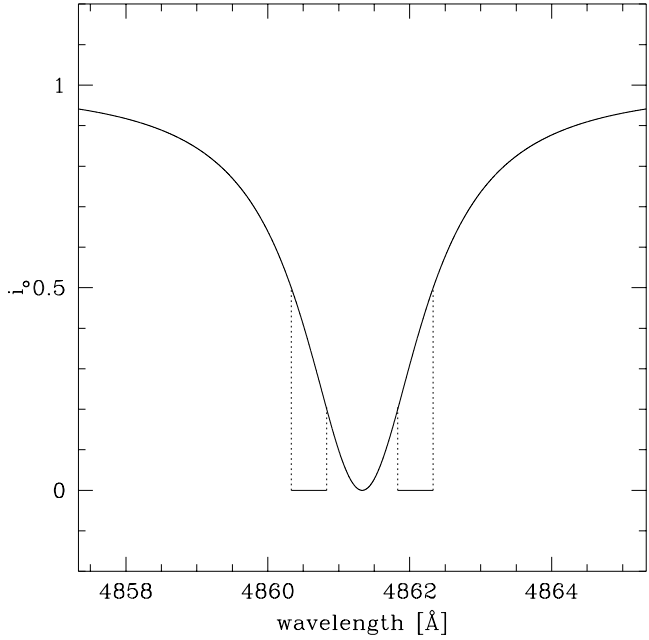}
\caption[]{The limits fixing the spectral range on which the fit is made are
randomly chosen on two segments symmetrically placed with respect to the centre
of the line.}
\end{figure}

\begin{figure}
\epsfysize=9cm
\leavevmode\epsffile{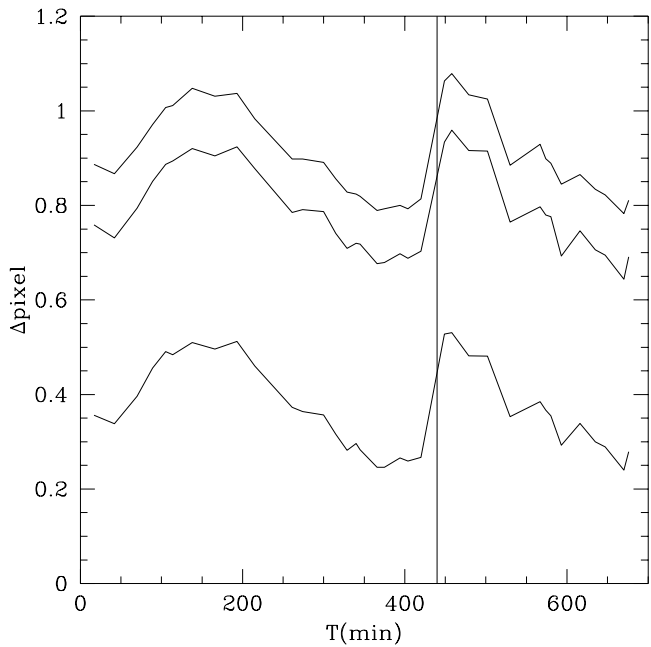}
\caption[]{Variation of the position of three lines of thorium during 6 and 7
November 1994. The vertical shift is arbitrary. The vertical line corresponds
to the filling of nitrogen.}
\end{figure}

\begin{figure}
\epsfysize=9cm
\leavevmode\epsffile{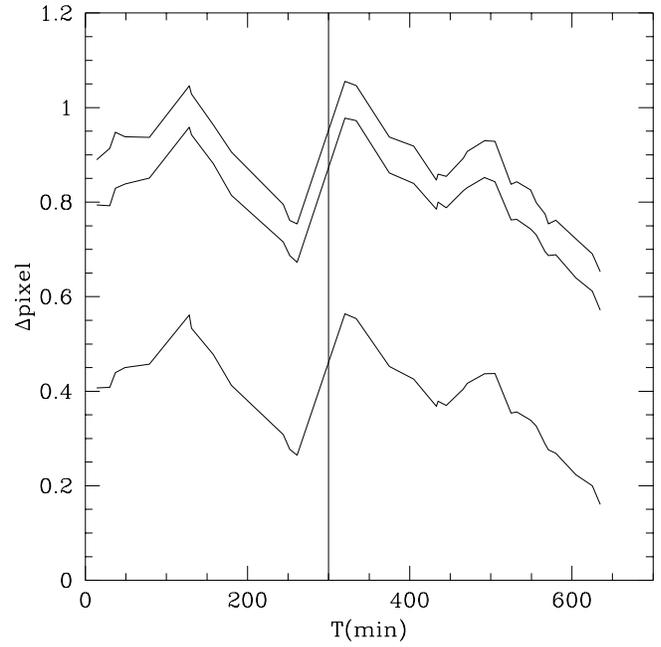}
\caption[]{Same as Figure 2, but for the night of 5 and 6 December 1994.}
\end{figure}

\subsection{Standard stars}
For standard stars, we prefer to use Coravel values in the system of faint IAU 
standard stars (Mayor \& Maurice 1985) than IAU values listed in the 
Astronomical Almanach because the latter are taken from various authors and 
sometimes not updated. For example, the IAU radial velocity for HD 114762
has been listed as 49.9$\pm$0.5 $kms^{-1}$ in the Astronomical Almanach since
at least 1981, while combined data from the Cfa and Coravel give a systemic
velocity of 49.35$\pm$0.04 $kms^{-1}$ (Latham et al. 1989).

For the four observing runs, Figure 4 shows the difference between radial 
velocities measured with Aur\'elie and the  Coravel value for each 
standard star. Our values are higher than the Coravel ones by about
2.75 km\,s$^{-1}$. 
The error bars are defined by the quadratic sum of the internal error (see 3.1.)
and the dispersion on the Coravel values. The internal precision of our 
measurements is very impressive: we obtain a scatter of about 0.4 km\,s$^{-1}$ 
around the mean radial velocity.

\begin{figure}
\epsfysize=9cm
\leavevmode\epsffile{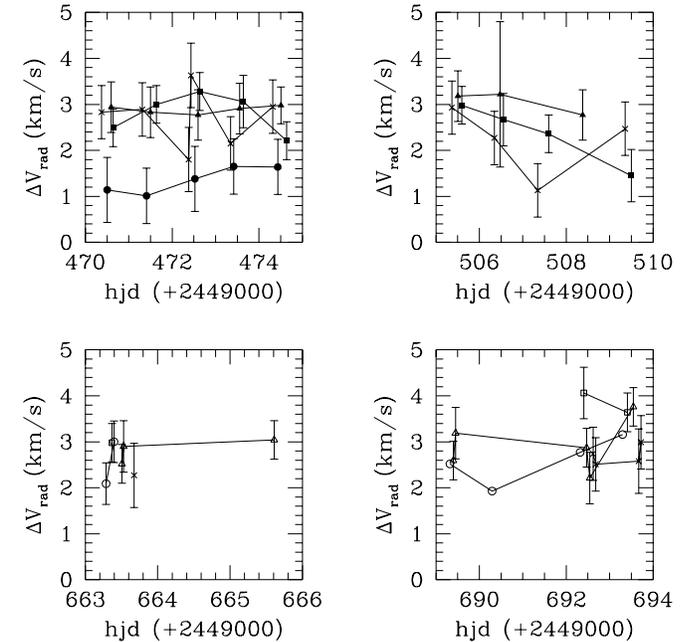}
\caption[]{Difference between Aur\'elie radial velocities and Coravel velocities
for the runs of May, June, November and December 1994. HD 693: open squares.
HD 22484: open triangles. HD 89449: crosses. HD 114762: full circles.
HD 136202: full triangles. HD 222368: open circles.}
\end{figure}

Notice that $\Delta V_{r}$ is clearly smaller for HD 114762 by about
1.4 km\,s$^{-1}$.
Such a difference can partly be due to the nature of this star, which has a very
small amplitude (Latham et al 1989 and Cochran et al 1991) and to the fact that 
our measurements were made precisely when the radial velocity 
was minimum: in this way we can explain a shift of about 0.6 km\,s$^{-1}$ with 
respect to the other standard stars. Unfortunately, we do not find any 
explanation for the remaining shift of 0.8 km\,s$^{-1}$. The reduction of the 
spectra was made again with MIDAS software,  only to find the same shift. So, we
cannot question the reduction.

We have simply subtracted the mean $\Delta V_{r}$ from the measured velocities.
The average $\Delta V_{r}$ are 2.80, 2.49, 2.69 and 2.90 km\,s$^{-1}$ for the 
runs of May, June, November and December respectively. For the first run,
the values of HD 114762 are not taken into account in the average value. Table 2
gives the individual corrected radial velocities and the mean velocity for 
standard stars; Table 3 lists the individual corrected radial 
velocities for the 50 giant A and F stars. 

\section{Binarity among the sample}
\subsection{Criterion of variability}

To take into account the errors of measurements on the determination of the
duplicity, we computed the $\chi^2$ value for each star of the programme:
$\chi^2=(n-1)(\frac{E}{I})^2$ where $n$ is the number of measurements, $E$ the
external error and $I$ the internal error. We then used an F-test which gives
the probability $P(\chi^2)$ that the variations of velocity are only due to the
internal dispersion. A star will be considered as double or intrinsically
variable if $P(\chi^2)$ is less than 0.01 (Duquennoy \& Mayor 1991).
Naturally, this test cannot say anything about the nature of the variability.

The distribution of $P(\chi^2)$ for non-variable stars should be flat from 0 to 
1, while the variable stars should gather at the smallest values of $P(\chi^2)$.
Therefore, this method allows to appreciate {\it a posteriori} the estimate of 
internal errors. Indeed, if these errors are underestimated, a gradient appears
in the distribution of $P(\chi^2)$ in favour of small values, while if they are 
overestimated, a peak appears near 1, indicating an abnormally strong 
predominance of constant stars.

\begin{figure}
\epsfysize=9cm
\leavevmode\epsffile{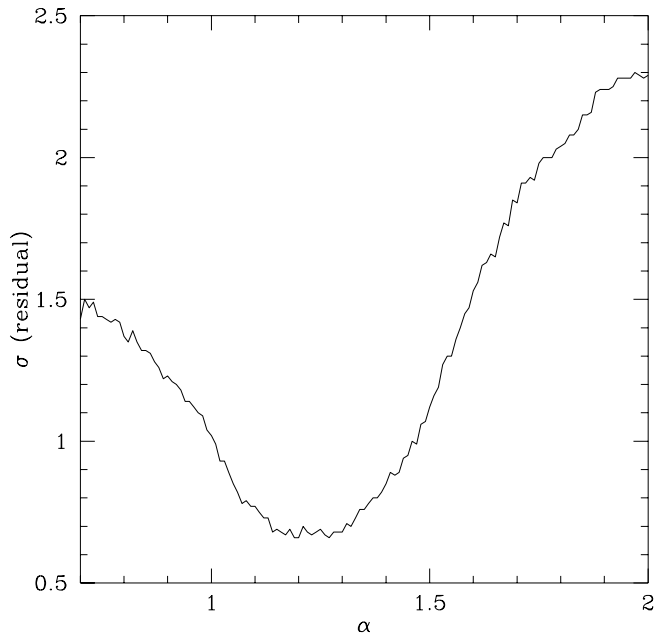}
\caption[]{Standard deviation of the residuals as a function of the $\alpha$ 
parameter.}
\end{figure}

\begin{figure}
\epsfysize=9cm
\leavevmode\epsffile{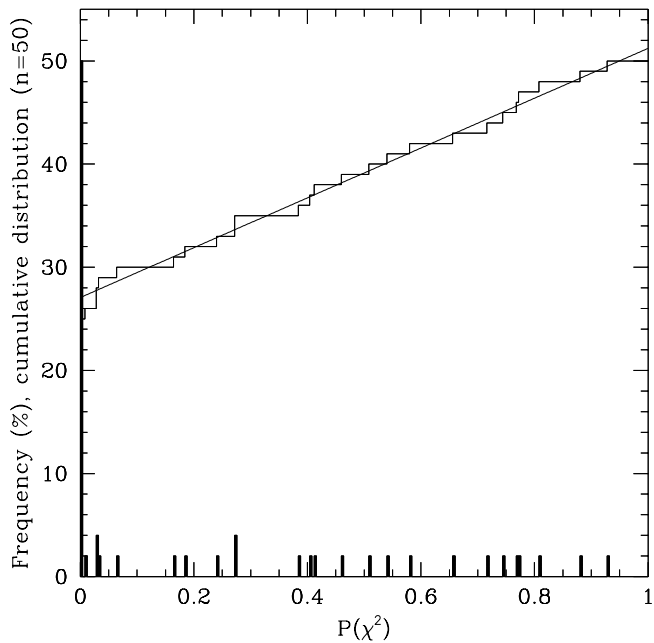}
\caption[]{Histogram and cumulative distribution of the $P(\chi^2)$ for the 
50 A and F giant stars. These distributions correspond to an $\alpha$ 
parameter of 1.2 ensuring a flat distribution between 0 and 1. We have also
drawn the straight line determined by a least-squares fit of the cumulative 
distribution.}
\end{figure}

In our case, the internal error is estimated by the quadratic sum of a term 
which depends on the width of the H$\beta$ line and the quality of the spectrum,
and a second term related with the instrumental shift during the night. While
the last term is rather well controlled, the first one is not well-known. 
Indeed, it strongly depends on the choice of the limits (see 3.2).
The internal error can be written:
\begin{equation}
I=\sqrt{(\alpha\sigma_{V_{r}}^s)^2+(\sigma_{V_{r}}^t)^2}
\end{equation}
where $\sigma_{V_{r}}^s$ is the dispersion due to $T_{\mathrm{eff}}$, $v\sin i$
and $S/N$,  $\sigma_{V_{r}}^t$ is the dispersion due to instrumental drift and 
$\alpha$ is adjusted to obtain a flat distribution of $P(\chi^2)$ on the 
interval [0,1], except for the small values of course. $\alpha$ indicates the 
quality of the preliminary estimation of $I$. To determine $\alpha$
quantitatively, one uses the cumulative distribution of $P(\chi^2)$ which must
approximate a straight line in the case of a flat distribution. For a given 
$\alpha$, one can compute the residuals to the regression line fitting the 
cumulative distribution. The $\alpha$ parameter corresponding to the minimum 
residuals is
then adopted. Figure 5 shows the behaviour of the r.m.s. deviation of the
residuals as a function of $\alpha$. A minimum clearly appears around
$\alpha = 1.2$. This means that 
the error on $\sigma_{V_{r}}^s$ is underestimated by about 20 \%, which is
quite reasonable considering the numerous uncertainties affecting its 
determination. Figure 6 shows the histogram and cumulative distribution of
$P(\chi^2)$ for $\alpha = 1.2$, as well as the straight line minimising the 
residuals of the cumulative distribution.

According to this criterion, 52 \% of the stars are variable and are listed in 
Table 4; the others are listed in Table 5. Each table gives the spectral type, 
$P(\chi^2)$, the blanketing parameter $\Delta m_2$, $v\sin i$ and eventually 
some remarks. The source of the projected rotational velocities is Abt \& 
Morrell (1995) or the Bright Star Catalogue (BSC), except for HD 6706,
HD 122703, HD 150453, HD 190172 and HD 217131 whose $v\sin i$ is determined by 
the optimum fit of a synthetic spectrum to our observed spectra. The spectral 
types are taken from Hauck (1986) who refers to Cowley et al (1969),
Cowley A.P. (1976), the Michigan catalogue (Houk and Cowley, 1975; Houk, 1978,
1982), Jaschek M. (1978) and the BSC. $\Delta m_2$ are taken from the Geneva
photometry database. The values can differ from those of Hauck (1986) because
new measurements have been made and a new reference sequence   for the Hyades
has been defined (see section 5.1). The $\Delta m_2$ value can be weaker by a
few thousandths of magnitude in the most unfavourable cases for visual doubles.
This effect can only diminish the sample of metallic F giants, while the sample
of metallic giants cannot be polluted by non-metallic stars. Remarks D and SB
come from the BSC.

The stars HD 2628 (3 measurements), HD 10845 (2), HD 11522 (2), HD 24832 (3),
HD 62437 (4), HD 69997 (4), HD 1772392 (10) and HD 187764 (7) belong to the 
catalogue of $\delta$ Scuti stars of Rodriguez et al (1994). In principle, all 
of these stars should be detected as variable, but the first three are not. For
these, we have only a few measurements separated by several days. As ill luck 
would have it, for HD 2628 and HD 11522 the exposures are made at the same
pulsational phase. For HD 10845, our measurements cover different phases, but
the small amplitude of the lightcurve (0.02 mag in the V filter) is probably
responsible for the non-detection. For 
the five $\delta$ Scuti stars detected, we find an average ratio of
110 km\,s$^{-1}$\,mag$^{-1}$ between the peak-to-peak radial velocity
and photometric variations, which is compatible with the value of 92
km\,s$^{-1}$\,mag$^{-1}$ given by Breger (1979). Therefore, it seems that the
$V_r$ variation of these five objects is only due to pulsation and not to any 
orbital motion.

Among the stars not detected as variable, five are listed as SB in the BSC:
HD 50019, HD 84607, HD 86611, HD 89025 and HD 92787. Low spectroscopic
dispersion (30-40 \AA\,mm$^{-1}$) and fast rotational velocity may probably 
explain the large variations reported in the past. Figure 7 shows for these
five stars the r.m.s scatter of the radial velocities in the literature as a
function of $v\sin i$. For HD 89025, we did not take into account the
measurements made by Henroteau (1923), because 
they differ systematically from the others and would generate an artificially
larger dispersion. For the older measurements, there is a clear correlation
between dispersion and rotation: when $v\sin i$ increases from 70 to
215 km\,s$^{-1}$, $\sigma$ increases from 
7 to 25 km\,s$^{-1}$. Our mean radial velocities values are compatible with
the older ones, except for HD 
86611 which rotates very fast.

\begin{figure}
\epsfysize=9cm
\centerline{\hbox{\psfig{file=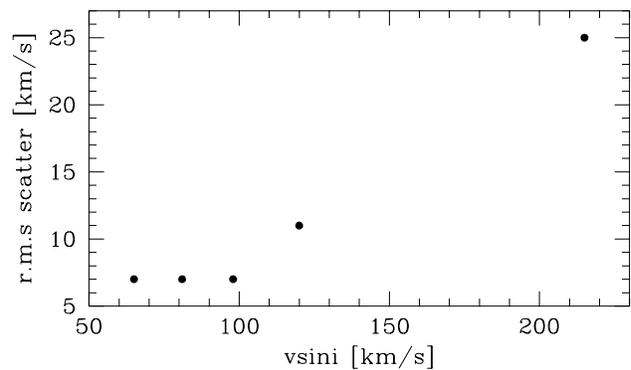,width=85mm,bbllx=20pt,bblly=210pt,bburx=550pt,bbury=524pt}}}
\caption[]{R.m.s scatter of the $V_r$ as a function of $v\sin i$ for stars
considered as SB in the BSC but not detected variable in this paper.}
\end{figure}

\begin{table*}
\hspace{6.9cm}{{\bf Table 4.} Detected variable stars.}
\vspace{2mm}   
\begin{center}
\begin{tabular}{r c r r r l}
\hline\multicolumn{1}{c}{HD}
     &\multicolumn{1}{c}{Spectral type} 
     &\multicolumn{1}{c}{$P(\chi^2)$} 
     &\multicolumn{1}{c}{$\Delta m_2$} 
     &\multicolumn{1}{c}{$v\sin i$} 
     &\multicolumn{1}{c}{Remarks} \\
\hline
4338 & F2III&0.000&0.009&98&D\\
24832 & F1V&0.000&0.004&150&$\delta$ Scuti\\
30020 & F4IIIp&0.000&0.089&60&SB\\
34045 & F2III&0.001&0.021&67&\\
60489 & A7III&0.000&0.008&15&\\
62437 & F0III&0.000&0.008&35&$\delta$ Scuti\\
69997 & F2III&0.000&0.042&25&$\delta$ Scuti\\
82043 & F0III&0.000&0.003&51\\
100418 & F9III&0.000&-0.020&33\\
103313 & F0V&0.000&0.009&61\\
104827 & F0IV-V&0.000&0.002&38&SB,D\\
118295 & A7-F0V&0.000&-0.003&135\\
122703 & F5III&0.000&0.008&69\\
150453 & F4III-IV&0.001&-0.033&10\\
155646 & F6III&0.000&-0.034&$\leq$10\\
159561 & A5III&0.000&-0.019&210&SB\\
171856 & A8IIIn&0.000&-0.001&110&D\\
174866 & A7Vn&0.001&-0.012&150&\\
176971 & A4V&0.000&-0.016&125&\\
177392 & F2III&0.000&0.030&120&$\delta$ Scuti\\
186005 & F1III&0.003&0.006&140&SB\\
187764 & F0III&0.000&-0.003&85&$\delta$ Scuti\\
190172 & F4III&0.000&-0.001&25\\
203842 & F5III&0.000&-0.006&84\\
209166 & F4III&0.001&0.007&$<$20&D\\
216701 & A7III&0.000&-0.005&80\\
\hline
\end{tabular}
\end{center}
\end{table*}

\begin{table*}
\hspace{6.9cm}{\bf Table 5.} Non-variable stars.
\vspace{2mm}
\begin{center}
\begin{tabular}{r c r r r l}
\hline\multicolumn{1}{c}{HD}
     &\multicolumn{1}{c}{Spectral type} 
     &\multicolumn{1}{c}{$P(\chi^2)$} 
     &\multicolumn{1}{c}{$\Delta m_2$} 
     &\multicolumn{1}{c}{$v\sin i$} 
     &\multicolumn{1}{c}{Remarks} \\
\hline
1671 & F5III&0.929&0.002&41\\
2628 & A7III&0.510&-0.004&18&$\delta$ Scuti\\
6706 & F5III&0.746&-0.003&50\\
10845 & A9III&0.770&0.015&85&$\delta$ Scuti\\
11522 & F0III&0.581&-0.009&120&$\delta$ Scuti\\
12573 & A5III&0.275&-0.017&95&\\
17584 & F2III&0.656&0.015&149\\
17918 & F5III&0.461&0.033&120\\
21770 & F4III&0.808&-0.023&29\\
48737 & F5III&0.065&0.008&70\\
50019 & A3III&0.275&-0.006&120&SB\\
84607 & F4III&0.388&0.023&98&SB\\
86611 & F0V&0.415&-0.015&215&SB\\
89025 & F0III&0.242&0.038&81&SB\\
92787 & F5III&0.719&-0.023&65&SB\\
108382 & A4V&0.881&0.006&65\\
150557 & F2III-IV&0.031&-0.009&67\\
178187 & A4III&0.032&-0.003&35\\
204577 & F3III&0.036&0.008&$\leq$15\\
205852 & F3III&0.775&0.020&155\\
210516 & A3III&0.407&-0.017&40\\
217131 & F3III&0.166&-0.016&66\\
219891 & A5Vn&0.186&-0.007&175\\
224995 & A6V&0.543&-0.009&90\\
\hline
\end{tabular}
\end{center}
\end{table*}

In Figure 8, we show the behaviour of the external scatter $E$ (which is
equivalent to the dispersion of the measurements) as a function of $v\sin i$
for the fifty giant stars and the seven standards of the programme. Black and
open symbols represent respectively non variable and variable stars on the basis
of the $P(\chi^2)$. A linear regression including only non-variable stars is
also represented. This straight line is, as a first approximation, the mean
internal error $I$ as a function of $v\sin i$ and agrees well with the values
determined previously. Most of the variable stars clearly appear above this
line and then we could also use it as criterion of variability.

\begin{figure}
\epsfysize=9cm
\leavevmode\epsffile{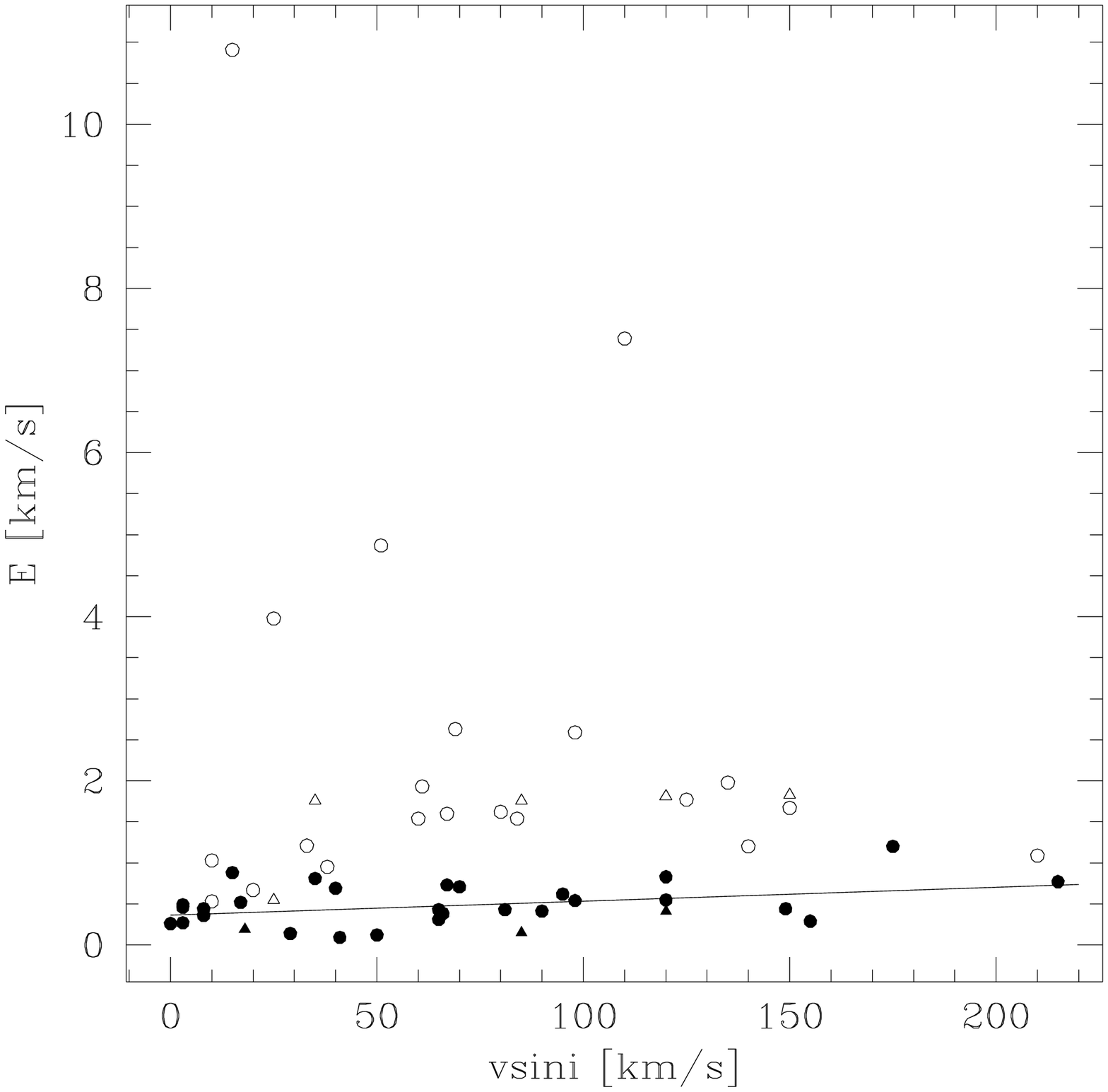}
\caption[]{External scatter $E$ as a function of $v\sin i$. Stars with
$P(\chi^2) > 0.01$ are represeted by black symbols and with
$P(\chi^2) \leq 0.01$ by open symbols. $\delta$ Scuti type stars are
represented by triangles. A linear regression is shown for constant stars.}
\end{figure}

$\delta$-Scuti type stars are represented by triangles, the most variable of
them having an external scatter of about 1.8 $km\ s^{-1}$, which is reasonable
for stars with an amplitude of 0.05 mag. We can see that most of the variable
stars have an external scatter below 1.8 $km\ s^{-1}$ and so the origin of this
variability remains ambiguous. Some of them are intrinsic variables not as yet
classified $\delta$ Scuti. Only stars with $E \geq 2 km\ s^{-1}$ can be
considered as binaries with a high probability.

\subsection{Rate of detection}

We have made a simulation to determine the rate of detected variable stars as a 
function of the period. For this, a sample of 1000 double stars with given 
periods was created as a first step. A flat distribution of the mass ratio
was assumed (Mazeh et al. 1992) with primary components of A and F types
(1.5-3$M_\odot$). The orbital elements $T_{\mathrm o}$, $\omega$ and $i$ are
randomly distributed, while the eccentricity is distributed according to 
Duquennoy \& Mayor (1991): when the period is less than 10 days, the orbit is
assumed to be circular; for  periods between 10 and 1000 days, the eccentricity
is distributed following a gaussian with a mean equal to 0.3 and $\sigma$=0.15
(cases with negative eccentricity were dropped and replaced); for longer 
periods, the distribution $f(e)=2e$ is assumed. In a second step, the radial
velocities of the created sample are computed at the epochs of observation 
of the real programme stars. Then a random internal error is added to these  
50000 radial velocities. Finally the $P(\chi^2)$ value of each star
is computed and we can take the census of detected binary stars for a given
period. The results are presented in Figure 9 for periods between 1 and $10^5$
days.

\begin{figure}
\epsfysize=9cm
\leavevmode\epsffile{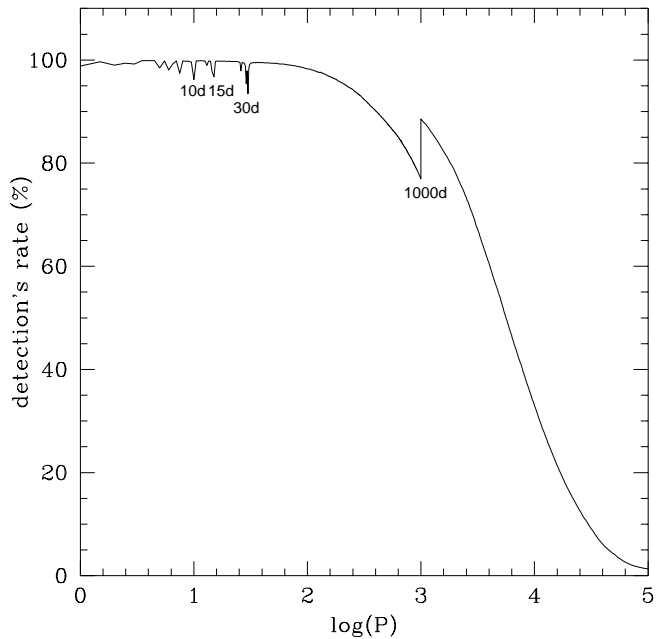}
\caption[]{Detection rate as a function of the period. $P$ is
given in days.}
\end{figure}

The simulated detection rate is very high for systems with periods below 100 
days: it varies between 93.5 \% and 99.9 \%. Above 100 days, the rate decreases 
rapidly, being about 80 \% for 1000 days and 30 \% for 10000 days. The 
discontinuity which appears for 1000 days is due to the strong change in the 
distribution of eccentricities: indeed, from this point on we grant more 
importance to large $e$. The size of this effect is related to the  time 
distribution of the measurements. For instance, if the exposures were more
distant in time, the discontinuity would be smaller. The simulation shows some 
very peaked depressions at shorter periods: at $P=30$ days, which corresponds
to the time interval between two successive observing runs and at dividers of
30, i.e. 15, 10, 7.5, 6 and 5 days. This is completely normal, because for such 
periods, the time distribution of the measurements makes the detection of
binary stars less efficient. At $P=3$ days and  $P=2$ days, the rate remains 
very high, because each run lasts for about 4 or 5 days. At $P=10$ days, there 
is a weak discontinuity due to the change of distribution of eccentricity, but 
this effect is hidden inside the peak.

In addition, we have computed the mean rate of detection among binaries with 
periods less than 100 days. The binaries are created as before but the periods 
are distributed as a gaussian with a mean equal to  
$\overline{log(P)}=4.8$ and $\sigma_{log(P)} = 2.3$ (Duquennoy \& 
Mayor 1991), where $P$ is given in days. When cut-offs at 1 and 100 days were
imposed, the detection rate reached 99 \%, i.e. all close binaries are detected.
The rate remains as high as 94 \% for periods between 1 and 1000 days.

\section{Comparison between Am stars and F giants}

Before discussing the rotation and the rate of binaries
 among Am and giant A-F stars, it is 
useful to consider again the mean value of the blanketing parameter $\Delta m_2$
for Am stars as well as the range of spectral types of giant metallic stars.

\subsection{Parameter of metallicity $\Delta m_2$}

In the $B2-V1$ vs $m_2$ diagram, the Am stars are located above the reference 
sequence defined by the Hyades (Hauck 1973). The difference $\Delta m_2$ is
interpreted in terms of metallicity. Hauck \& 
Curchod (1980) have determined the mean value of $\Delta m_2$ for classical Am 
stars ($g\geq 5$, where $g$ is the difference between the spectral type deduced
from the metallic lines and that from the K line) and for mild Am stars
($g<5$) and have obtained respectively 0.013 $\pm$ 0.019 (146 stars) and 0.002
$\pm$ 0.011 (23 stars). They considered only objects brighter than the 7th
magnitude  in order to avoid any significant interstellar reddening.

Since then, a new sequence of reference has been defined (Hauck et al. 1991)
taking into account new observations of the Hyades. This new sequence is
different from the old one mainly for stars with a type later than F; for early 
A type stars the old and new sequences are identical, while for late A and F
stars they are separated by only a few thousandths of a magnitude. Taking into 
account the stars of the revised catalogue of Curchod \& Hauck (1979) brighter
than the 7th magnitude, the new values are 
$\overline{\Delta m_2} = 0.011 \pm 0.021$ (238 stars) and  $\overline{\Delta 
m_2} = -0.001 \pm 0.025$ (101 stars) for classical and mild Am respectively,
i.e a decrease of about 2 or 3 mmag with respect to the preceding values.
Therefore, a giant star will be considered here as metallic whenever its
$\Delta m_2$ is larger than or equal to 0.013, rather than 0.015 as defined by 
Hauck \& Curchod (1980). 

\subsection{Spectral types of metallic giants}

The diagram $\Delta m_2$ vs $T_{\mathrm{eff}}$ for giant A-F stars of Hauck
(1986) is shown again in Figure 10. The effective temperature is deduced from
the semi-empirical calibrations of K\"unzli et al (1997). Notice that only F
stars have an enhanced $\Delta m_2$ value while this property never applies for
A-type stars. Indeed, all stars with $\Delta m_2 \geq 0.013$ have types
between F0 and F6, except for HD 10845, HD 90277 and HD 147547 which are
classified A9 and HD 4849 which is classified A9/F0.
The lack of metallic stars later than F6 is explained by the diffusion theory: 
Vauclair \& Vauclair (1982) have defined the limit where the diffusion time
scale for helium at the bottom of the surface convective zone equals the
stellar lifetime in a $log(L/L_{\odot})$ vs $log(T_{\mathrm eff})$ diagram;
this limit crosses the area of giants at $log(T_{\mathrm{eff}})=3.8$, which fits
exactly the limit we find in our diagram. It is most interesting to notice that
several metallic giants also are $\delta$ Scuti stars, because Am peculiarity
and $\delta$ Scuti-type pulsation are mutually exclusive. Only mild Am stars
may be $\delta$ Scuti, as well as $\delta$ Del stars. In this respect, the
metallic F giants are completely similar to $\delta$ Del stars (Kurtz 1976).

The study of a relation between Am stars and metallic A-F giants is now 
restricted between Am and metallic giant F0-6 stars. In the next two section,
we compare rotation and duplicity among these two samples.

\begin{figure}
\epsfysize=9cm
\leavevmode\epsffile{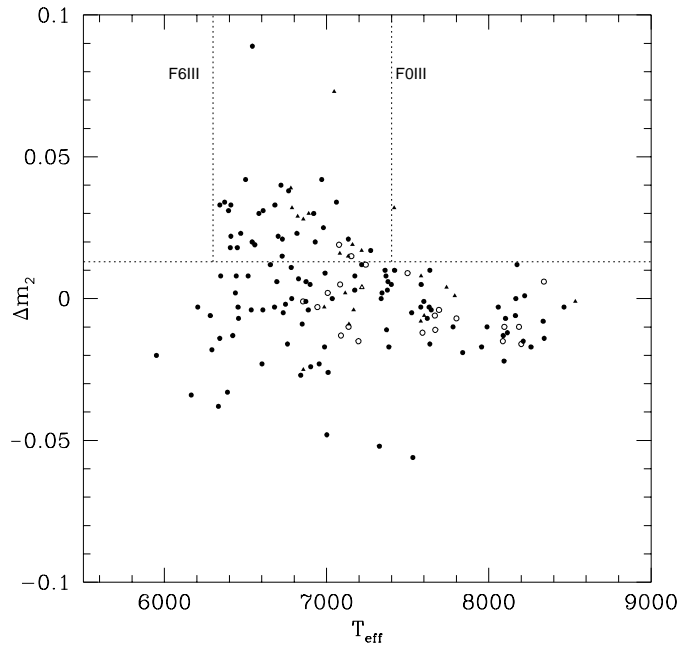}
\caption[]{$\Delta m_2$ vs $T_{\mathrm{eff}}$ diagram for A-F giants of
Hauck (1986). The black dots represent spectroscopic giant stars and open dots
photometric giants ($\Delta d \geq 0.12$). $\delta$ Scuti stars are represented
by triangles, the 
others by circles. The horizontal line separates metallic stars from normal 
ones. The two vertical lines at 6300 K and 7400 K define the red and blue limits
for metallic giants.}
\end{figure}

\subsection{$V\sin i$ of Am and metallic F0-6III}

The Am stars are taken from the revised catalogue of 
Curchod \& Hauck (1979). Their $v\sin i$ used is an average of the values given
by Abt \& Levy (1985), Uesugi \& Fukuda (1978,1982), Bernacca \& Perinotto
(1971), Boyarchuk \& Kopylov (1964). For stars with four measurements, we obtain
a standard deviation of about 9 $kms^{-1}$ which is a first indication of the
precision of these $v\sin i$. The projected rotational velocities of metallic
F0-6III stars are taken from Abt \& Morrell (1995) or the BSC (28 stars).
Abt \& Morrell (1995) give an error on the determination of the radial velocity
of about 10 $kms^{-1}$. To make the comparison meaningful,
we take into account only the Am stars brighter than the 7th magnitude and with 
$\Delta m_2 \geq 0.013$ (98 stars). The Am stars with a low $\Delta m_2$ value
have underabundant scandium and calcium, and normal or slightly overabundant
heavier elements. On the other hand, F0-6III stars with low $\Delta m_2$ have a 
normal chemical composition, so there is no similarity between these two
categories.

We present in Figure 11 the histograms and cumulative distributions of the 
projected rotational velocities for the Am and metallic F0-6III stars, and the
cumulative distribution for non metallic FIII stars. Obviously,
Am stars (full line) and metallic F0-6 giants (dotted line) do not follow the 
same distribution. The $v\sin i$ of Am stars are below 100 km\,s$^{-1}$ except 
for one star, while those of F giants are often faster than 100 km\,s$^{-1}$.
The maximum of the  distribution for Am star is between 30 and 40 km\,s$^{-1}$; 
for F giants, the distribution seems flat with a cut-off at about 160 
km\,s$^{-1}$.

\begin{figure}
\epsfysize=9cm
\leavevmode\epsffile{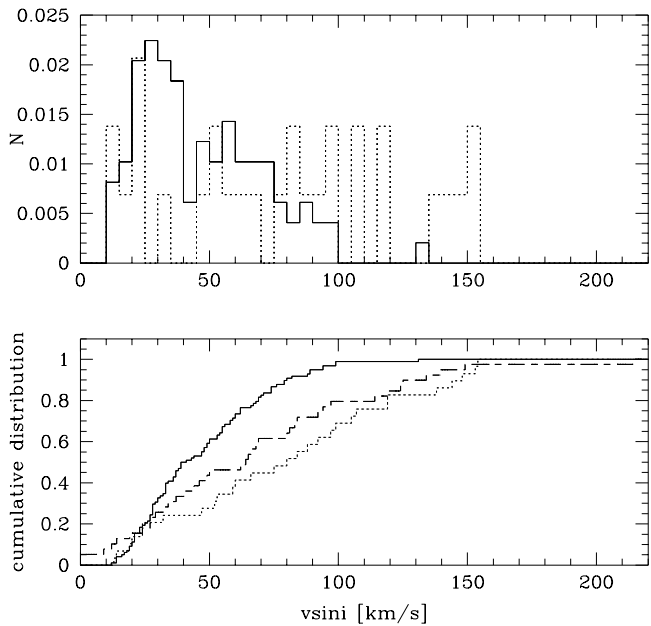}
\caption[]{Comparison between the distributions of $v\sin i$ for Am stars
(full line) and metallic F0-6III stars (dotted line). We have also given the
cumulative distribution for giant non-metallic F stars (broken line). Histograms
and cumulative distributions are normalised.}
\end{figure}

Cumulative distributions give more satisfactory information than histograms 
because they do not depend on the width of the bins. We have applied the
Kolmogorov-Smirnov test to the distributions of the Am stars and of the metallic
giants: adopting the $H_o$ hypothesis 
that they are identical, one obtains the probability $P = 3.692\ 10^{-4}$ to
have a test value at least as extreme as the actually observed one. Therefore
we can reject the $H_o$ hypothesis at the 99 \% confidence level and the
distributions are very probably not similar. In the sample of F giants, some
$\delta$ Scuti stars are present; in the framework of the scenario where Am
stars are progenitors of metallic F giants, this implies  that Am stars may
pulsate as soon as they have evolved into giants. This assumption is based
on the fact that some $\delta$ Del stars  do pulsate and may be evolved Am
stars (Kurtz 1976). If this assumption is wrong, then the sample of metallic
giants is polluted. For that reason, we have applied the test used above to
Am and F giants not known as $\delta$ Scuti stars and found
$P = 6.227\ 10^{-4}$, which does not change the conclusion.

The Kolmogorov-Smirnov test applied to metallic and non-metallic FIII stars
gives a $P$ value of 0.46. So there is no real difference between these two
samples from the point of view of projected rotational velocities, contrary
to what we observe between Am and normal A dwarf.

\subsection{Rate of binaries among Am and metallic F0-6III}

Let us discuss now the rate of close binaries with periods below 1000 days
among  metallic F0-6 giants.

Since there are only 10 such stars in our sample of stars measured with Aurélie,
we have to rely largely on results published in the literature. Table 6 lists
some remarks about these data for each of the 41 metallic giants. For HD 4919
and HD 177482, however, the $V_r$ data are too poor and inaccurate. Only 39
stars have therefore been retained, of which 6 stars are  strictly speaking
members of tight binaries: HD 30020,
HD 34045, HD 43905, HD 85040, HD 108722 and HD 110318, which constitute
15.4 \% of the sample. For some other stars, the decision is less clear-cut.
For fast rotators, it is difficult 
to know whether the observed dispersion is just due to the large
$v\sin i$ or betrays an orbital motion; this is especially the case of HD 13174 
and HD 147547. For the $\delta$ Scuti stars HD 214441, we observe a ratio of
about 200 $kms^{-1}mag^{-1}$ between the peak-to-peak radial velocity and
photometric variation which is double the value given by Breger (1979). Thus
the additional variation in radial velocities could be due to  orbital motion.
But if these three stars are added to the sample of tight 
binaries, the proportion would only increase to 23.1 \% and to 23.3 \% if we
take only F giants not known as $\delta$ Scuti type stars. These values are the
observed rates; the real rates must be only slightly higher because the fraction
 of undetected stars is weak for small periods. It is interesting to
note that the observed rate agrees with the value for solar-type stars in the
vicinity of the Sun, i.e. 21.7 \% (Duquennoy \& Mayor 1991) for $P \leq$ 1000
days.

In their paper of 1985, Abt \& Levy estimated that the number of Am stars in
double systems with periods less than 1000 days represents 75 \% of the sample, 
i.e. a rate considerably larger than what we find for metallic F giants.
This is an additional reason to reject the possibility of any evolutionary
link between Am stars and metallic F giants.

The upper limit of the fraction of binary stars among the forty  non-metallic
A and F giant stars measured at OHP is 47 \%. This value is enhanced because
it certainly includes some unrecognised intrinsic variables. Of the 113 stars
of this category (Hauck 1986), 24 are classified as SB, 11 have a variable
radial velocity (V) and 6 are suspected variables (V?) in the BSC. The fraction
of binaries with a small period among this type of stars is between 21.2 \%
(SB only) and 36.3 \% (SB+V+V?) which may include some intrinsic variables.
Thus frequencies of binaries with small periods among metallic and non-metallic
A and F giant stars are not so different from each other contrary to the case of
Am and normal A dwarfs.

\section{Conclusion}

From the point of view of observed chemical abundances and of the theory of
radiative diffusion, the scenario which considers Am stars as progenitors
of metallic F0-6 giants  is completely justified. It was worth the effort to
try to settle this idea on firmer grounds, or alternatively to question it,
by considering two other fundamental characteristics of Am stars, namely
slow rotation and high rate of binaries.
The main results of our work may be summarised as follows:
\begin{itemize}
\item[-] Giant A-type stars never show an enhanced photometric 
metallicity, i.e $\Delta m_2 \geq 0.013$. Therefore, if they are descendants
of Am stars, they have not retained their chemical peculiarity. Most
of them have probably never been Am in the past. 
\item[-]One-third of the metallic F0-6 giants are fast rotators with
$v\sin i > 100 kms^{-1}$, while practically all Am stars are slow rotators with
$v\sin i < 100 kms^{-1}$. Moreover, the shape of the
$v\sin i$ distribution is completely different in each case: for metallic
F0-6III stars it is flat with a cut-off at about 160 km\,s$^{-1}$, while that
of Am stars shows a maximum between 30 and 40 km\,s$^{-1}$ with a steady 
decrease towards 100 km\,s$^{-1}$. This still holds valid when the giant F
sample is restricted to stars not known as $\delta$ Scuti type.
\item[-]The non-metallic giants have the same $v\sin i$ distribution as the
metallic ones.
\item[-]The rate of binaries among normal A-F giants is no more than 47 \% for
orbital periods less than 1000 days. This fraction may be overestimated because
part of the $V_r$ variations observed might be due to $\delta$ Scuti-type
variations rather than orbital motion.
\item[-]For metallic F giants (considering or not $\delta$ Scuti stars), the
best data in the literature together with our measurements indicate a rate of
binaries with $P<1000$ days of less than 30 \%. This is smaller, though not
significantly so, than for normal giants. We may have missed binaries with a
fast rotating primary, since the best $V_r$ data in the literature concern
sharp-lines stars (hence slow rotators). However, one sees no reason why most
binaries should be in this case, especially as tidal friction would tend to
slow down axial rotation. For Am stars, this rate is 75 \% (Abt \& Levy 1985).
\end{itemize}

It seems  therefore difficult to admit that Am stars can be progenitors of
metallic F giants. If such was the case, one should indeed expect:
\begin{enumerate}
\item A larger rate of binaries among metallic giants than among normal ones,
its value being close to that of Am stars (75 \%)
\item A $v\sin i$ distribution of metallic giants strongly peaked at very small
values, with a tail extending to less than 100 kms$^{-1}$. Indeed, the giants
are  all about to leave the main sequence and have, on average, larger radii
than Am stars. Therefore they can only rotate more slowly, by conservation of
angular momentum.
\item Widely different $v\sin i$ distributions for the normal, than for the
metallic giants, as is the case for  the normal A dwarfs compared with the Am
stars.
\end{enumerate}

None of these three expectations is fulfilled. As a whole, the metallic giants
cannot be considered as evolved Am stars, although the previous considerations
do not exclude the possibility that some of them (especially the slower
rotators) may have been Am stars in the past. 

One might object  that even fast rotating metallic giants may have main sequence
metallic progenitors, but the latter have gone unnoticed by the classifiers
because of fast rotation. But such progenitors would have been detected by
Geneva or Str\"omgren photometry (through the $m_2$ or $m_1$ parameters), while
significant photometric metallicity is observed only in stars classified
spectroscopically as Am, except for the metallic F giants.
The assumption of overabundances remaining steady (apart from that of Ca) up to
the very end of the main sequence life therefore seems wrong, and diffusion
theory indeed predicts that they may change drastically on shorter timescales,
at least near the ZAMS (Alecian 1996).

But if metallic F giants are not evolved Am stars, what is their origin then?
The fact  that they are not especially slow rotators is intriguing, because it
suggests that no special initial conditions are required to produce them. The
same can be said about the rate of binaries
, which does not seem special either. In view of this, the very simplest
alternative to the idea of Am progenitors is to speculate that every late A and
early F star goes through a short phase of enhanced atmospheric metallicity
around the end of its life on the main sequence.

We have seen that for some FIII stars the metallicity can coexist with high
projected rotational velocity ($\geq 100 kms^{-1}$). The explanation of this
fact is a real challenge addressed to theoreticians of diffusion, because for
main sequence stars  metallicity can appear only in slow rotators
($\leq 100 kms^{-1}$).

\acknowledgements{This research has made use of the Simbad database, operated
at CDS, Strasbourg, France. We thank Drs. Michel Mayor and Stéphane Udry
(Geneva Observatory) for providing us Coravel data for standard stars.
We thank Mrs B. Wilhelm for the correction of the English text. This paper
received the support of the Swiss National Science Foundation.}

\appendix
\section{Importance of the choice of the limits for the fit of the $H\beta$
line}

Simulations were carried out to observe the behaviour of
$\overline{\lambda_0(H\beta)}$ and $\sigma_{\lambda_0(H\beta)}$ as a function
of the choice of the two segments in order to find their optimal position and
length.

Synthetic spectra were computed at 6000 K and 7500 K with a log$g$ of 4 and
solar metallicity. They were then convoluted by instrumental and rotational
profiles. We considered  projected rotational velocities of 50 and
100 km\,s$^{-1}$. Finally, a random noise was added, corresponding to a
signal-to-noise ratio
of 150.

Two types of simulations were performed:
\begin{enumerate}
\item Two internal limits of the segments were fixed symmetrically with
respect to the centre of the line, the external limits being moved  away
progressively from the line's centre. $\overline{\lambda_0(H\beta)}$ and
$\sigma_{\lambda_0(H\beta)}$ was then computed for each position of these
segments.

\item The external limits of the segments were fixed, the internal limits
being brought regularly towards the centre of the line and
$\overline{\lambda_0(H\beta)}$ and $\sigma_{\lambda_0(H\beta)}$ were again
computed.
\end{enumerate}

The results are given in Figures 12 and 13, the upper part of them showing the
variation of $\sigma_{\lambda_0(H\beta)}$ and the lower part the variation of
$\overline{\lambda_0(H\beta)}$ as a function of the limits for the 4 spectra.
Each curve corresponds to different initial conditions, i.e. to different fixed
limits. The first type of simulation is represented by dotted lines and the
second type by dots and dashes. The dispersion and the mean central wavelength
are given as a function of the moving limit. To make these graphics easier to 
read, we have represented the right wing of the $H\beta$ line with the scale
in normalised flux (right axis of the graphics).

First, let us discuss the behaviour of $\sigma_{\lambda_0(H\beta)}$. When the 
internal and external limits are the same on each wing, the dispersion is
evidently null, increasing when the internal and external limits are separated.
This increase depends on each profile as seen in Figures 12 and 13.
Nevertheless, we can make general remarks. First, the nearer the limits to the
centre of the line, the stronger the dispersion. This is easily understood,
because near the centre only a few points are taken into account to make the fit
and a weak change of limits leads to important changes of the three parameters
of $L(\lambda)$. These parameters are not well defined either when we choose
external limits too far 
from the centre, because $H\beta$ have a lorentz profile only near the centre. 
Thus limits that are neither too close nor too far from the centre have to be 
chosen. Generally speaking, these subjective criteria are satisfactory when 
the segments lie on the linear part of the $H\beta$ profile.

Increased $v\sin i$ and $T_{\mathrm{eff}}$ make  the determination of the
central wavelength more difficult because the line is wider. The dispersion
therefore increases with $v\sin i$ and $T_{\mathrm{eff}}$. When the segments
are on the linear part 
of the profile, the following dispersions result: at 6000 K,
$\sigma_{\lambda_0(H\beta)}$ varies from 0.003 to 0.007 \AA\ for $v\sin i$
equal to 50 and 100 km\,s$^{-1}$ respectively and at 7500 K, its values are
0.004 to 0.007 
\AA\ for the same $v\sin i$. Thus the dispersion mainly depends on the
rotational velocity and barely on the effective temperature.

\begin{figure}
\epsfysize=9cm
\leavevmode\epsffile{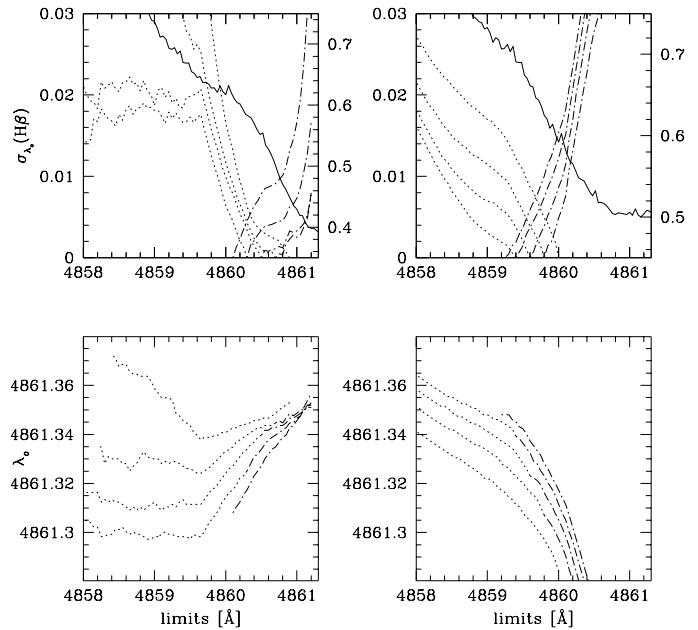}
\caption[]{Variation of $\overline{\lambda_0(H\beta)}$ and
$\sigma_{\lambda_0(H\beta)}$ as a function of the limits for a star of 6000 K  
with a log$g$ of 4 and a solar metallicity. The graph at the left simulates 
variations for a $v\sin i$ of 50 km\,s$^{-1}$ and at the right of
100 km\,s$^{-1}$. See text for 
comments.}

\end{figure}
\begin{figure}
\epsfysize=9cm
\leavevmode\epsffile{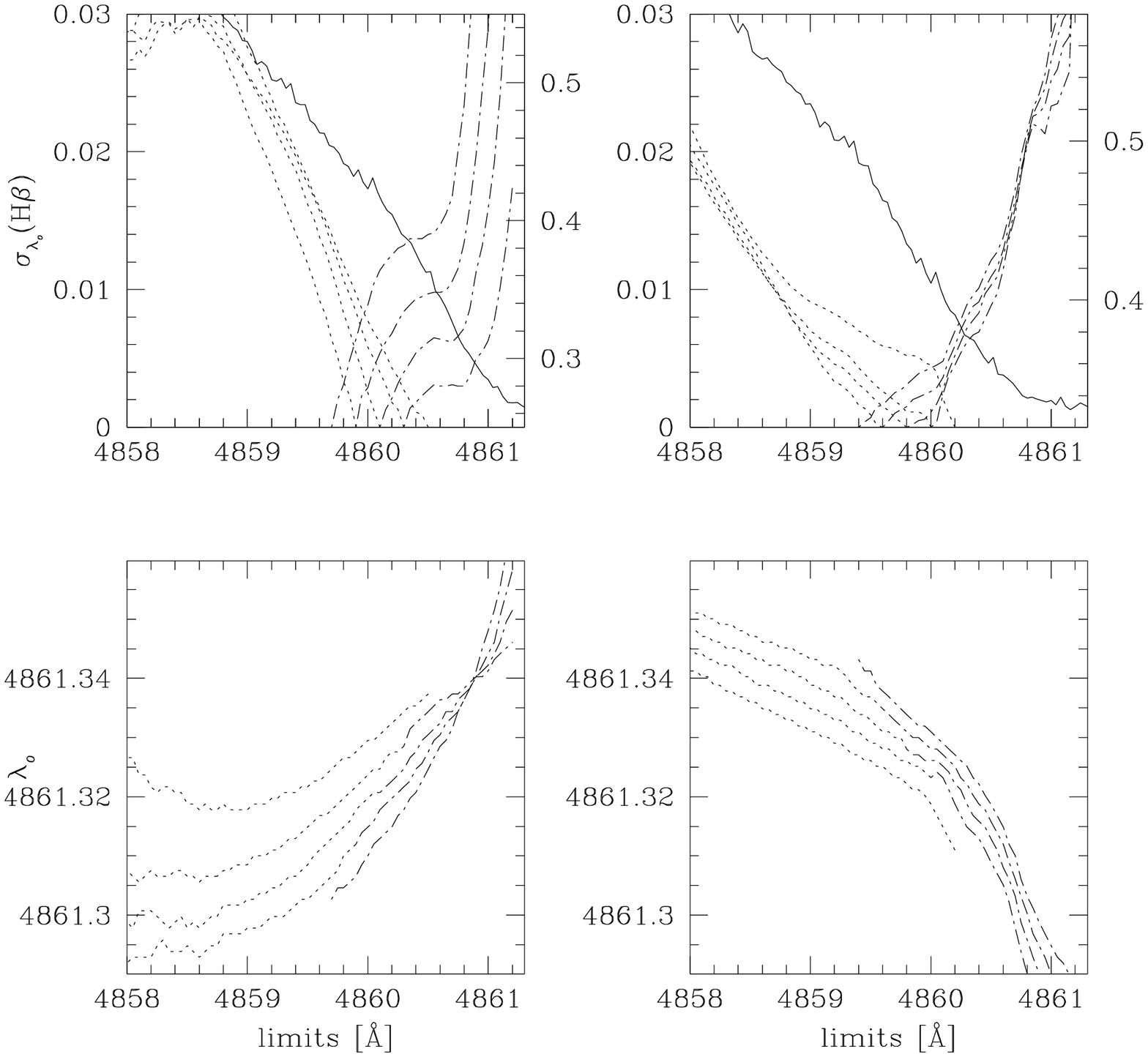}
\caption[]{Same as Figure 11, but with $T_{\mathrm{eff}}=7500 K$.}
\end{figure}

On the lower graphics, one sees that $\overline{\lambda_0(H\beta)}$ varies much 
as a function of the position and lengths of the segments. Nevertheless, if we
consider only the linear portion, the variability of this parameter is not so
important and 
corresponds to values given above. We observe that the shape of these 
fluctuations depends essentially on the $v\sin i$: for a $v\sin i$ of
100 km\,s$^{-1}$, 
$\overline{\lambda_0(H\beta)}$ regularly decreases when the limits approach the 
centre of the line; for a $v\sin i$ of 50 km\,s$^{-1}$ we observe an opposite
trend. 

The results of these simulations give only a qualitative idea about the 
behaviour of $\overline{\lambda_0(H\beta)}$ and $\sigma_{\lambda_0(H\beta)}$ as 
a function of the chosen segments. We have shown that these parameters
are very sensitive to the limits, which have to be put on the linear parts of 
the line profile to alleviate this problem. For a given star, we always use 
the same limits in order 
to have a good internal coherence. Nevertheless, we may have systematic errors, 
but this is not a severe problem because we are interested in variations of
radial velocity rather than in absolute values.

\vspace{0.5cm}

\begin{table}
\hspace{0cm}{\bf Table 2.} Radial velocities of standard stars
\begin{flushleft}
\begin{tabular}{rcrc}
\hline
HD  & HJD     & $V_r$ & $I$ \\
    & (+2449000)&km\,s$^{-1}$&km\,s$^{-1}$\\
\hline
693 & 663.367 & 15.12 &0.28\\
693 & 692.408 & 15.99 & 0.48\\
693 & 693.411 & 15.57 &0.28\\
    &    $\overline{V_{r}}=$     & 15.44$\pm$0.36&\\
\\
22484 & 663.507 & 27.59 &0.28\\
22484 & 663.528 & 27.97 & 0.48\\
22484 & 665.610 & 28.11 &0.28\\
22484 & 689.403 & 27.45 &0.28\\
22484 & 689.450 & 28.05 & 0.48\\
22484 & 692.473 & 27.73 &0.28\\
22484 & 692.544 & 27.07 & 0.48\\
22484 & 693.546 & 28.62 &0.28\\
    &   $\overline{V_{r}}=$      & 27.86$\pm$0.44&\\
\\
89449 & 470.375 & 6.07 &0.28\\
89449 & 471.311 & 6.13 &0.28\\
89449 & 472.383 & 5.04 & 0.48\\
89449 & 472.433 & 6.87 & 0.69\\
89449 & 473.347 & 5.39 &0.28\\
89449 & 474.319 & 6.19 &0.28\\
89449 & 505.366 & 6.49 &0.28\\
89449 & 506.346 & 5.83 &0.28\\
89449 & 507.345 & 4.69 &0.28\\
89449 & 509.356 & 6.03 &0.28\\
89449 & 663.674 &  5.62 & 0.48\\
89449 & 692.618 &  5.88 &0.28\\
89449 & 692.681 &  5.65 &0.28\\
89449 & 693.668 &  5.72 & 0.48\\
89449 & 693.722 &  6.13 &0.28\\
    &       $\overline{V_{r}}=$  & 5.84$\pm$0.52&\\
\\
114762 & 470.503 & 47.51 & 0.48\\
114762 & 471.411 & 47.38 &0.28\\
114762 & 472.526 & 47.75 & 0.48\\
114762 & 473.420 & 48.02 &0.28\\
114762 & 474.432 & 48.01 &0.28\\
    &      $\overline{V_{r}}=$   & 47.77$\pm$0.26&\\
\hline
\end{tabular}
\end{flushleft}
\end{table}

\begin{table}
\hspace{0cm}{\bf Table 2 (continued).} 
\begin{flushleft}
\begin{tabular}{rcrc}
\hline
HD  & HJD     & $V_r$ & $I$ \\
    & (+2449000)&km\,s$^{-1}$&km\,s$^{-1}$\\
\hline
136202 & 470.593 & 54.33 & 0.48\\
136202 & 471.499 & 54.22 & 0.48\\
136202 & 472.597 & 54.16 & 0.48\\
136202 & 473.555 & 54.30 & 0.48\\
136202 & 474.504 & 54.37 &0.28\\
136202 & 505.503 & 54.89 & 0.48\\
136202 & 506.474 & 54.93 & 1.56\\
136202 & 508.381 & 54.48 & 0.48\\
    &    $\overline{V_{r}}=$     & 54.39$\pm$0.27\\
\\
187691 & 470.642 &  -0.58 &0.28\\
187691 & 471.638 &  -0.07 &0.28\\
187691 & 472.638 &  \ 0.21 &0.28\\
187691 & 473.640 &  -0.01 & 0.48\\
187691 & 474.639 & -0.86 &0.28\\
187691 & 505.598 & \ 0.23 &0.28\\
187691 & 506.558 & -0.08 & 0.48\\
187691 & 507.599 & -0.39 &0.28\\
187691 & 509.501 & -1.30 & 0.48\\
    &   $\overline{V_{r}}=$      &  -0.27$\pm$0.49&\\
\\
222368 & 663.287 &  4.76 &0.28\\
222368 & 663.395 &  5.67 &0.28\\
222368 & 689.321 &  4.98 & 0.48\\
222368 & 690.296 &  4.39 &0.28\\
222368 & 692.320 &  5.23 &0.28\\
222368 & 693.300 &  5.62 &0.28\\
    &   $\overline{V_{r}}=$      & 5.12$\pm$0.45&\\
\hline
\end{tabular}
\end{flushleft}
\end{table}

\begin{table}
\hspace{0cm}{\bf Table 3.} Radial velocities of A and F giants
\begin{flushleft}
\begin{tabular}{rcrc}
\hline
HD  & HJD     & $V_r$ & $I$ \\
    & (+2449000)&km\,s$^{-1}$&km\,s$^{-1}$\\
\hline
1671 & 663.445 & 10.53 & 0.48\\
1671 & 689.354 & 10.55 & 0.28\\
1671 & 693.444 & 10.34 & 0.48\\
\\
2628 & 663.451 &  -10.86 &0.28\\
2628 & 689.366 &  -10.67 &1.12\\
2628 & 693.448 &  -10.40 &0.28\\
\\
4338 & 663.410 &  2.08 & 0.69\\
4338 & 689.381 & -0.47 & 1.34\\
4338 & 693.470 &  5.84 & 0.94\\
\\
6706 & 663.426 & 23.81 &0.28\\
6706 & 692.487 & 23.57 & 0.69\\
\hline
\end{tabular}
\end{flushleft}
\end{table}

\begin{table}
\hspace{0cm}{\bf Table 3 (continued).} 
\begin{flushleft}
\begin{tabular}{rcrc}
\hline
HD  & HJD     & $V_r$ & $I$ \\
    & (+2449000)&km\,s$^{-1}$&km\,s$^{-1}$\\
\hline

10845 & 663.458 &  -13.97 & 0.48\\
10845 & 689.412 &  -13.67 & 0.90\\
\\
11522 & 663.436 &  1.16 & 0.69\\
11522 & 693.514 &  1.99 & 1.34\\
\\
12573 & 663.471 &  8.62 & 0.69\\
12573 & 692.522 &  7.38 & 0.90\\
\\
17584 & 689.423 & 18.41 & 0.94\\
17584 & 663.481 & 18.60 & 0.90\\
17584 & 692.387 & 19.42 & 0.69\\
\\
17918 & 663.491 & 17.21 & 0.48\\
17918 & 663.552 & 18.40 & 0.69\\
17918 & 689.434 & 17.11 & 0.90\\
17918 & 692.396 & 18.08 & 0.90\\
\\
21770 & 663.500 &  -46.18 &0.28\\
21770 & 689.444 &  -46.35 & 0.48\\
21770 & 692.469 &  -46.01 &0.28\\
\\
24832 & 663.538 & 20.61 & 1.56\\
24832 & 692.433 & 18.20 & 0.90\\
24832 & 693.569 & 22.67 & 0.69\\
\\
30020 & 663.570 & 34.31 & 0.69\\
30020 & 692.567 & 37.19 & 0.48\\
30020 & 693.609 & 37.87 & 0.90\\
\\
34045 & 663.591 & 29.89 & 0.48\\
34045 & 689.467 & 26.19 & 0.94\\
34045 & 692.587 & 29.18 & 0.48\\
\\
48737 & 663.606 & 26.12 & 0.90\\
48737 & 665.620 & 26.30 & 0.48\\
48737 & 689.459 & 25.77 & 0.48\\
48737 & 692.598 & 27.35 & 0.69\\
48737 & 693.630 & 27.56 & 0.48\\
\\
50019 & 663.611 & 31.17 & 0.69\\
50019 & 664.643 & 30.75 & 0.69\\
50019 & 665.626 & 30.30 & 0.69\\
50019 & 692.601 & 30.04 & 0.69\\
50019 & 693.633 & 32.39 & 0.90\\
\\
60489 & 663.617 & 52.67 & 0.69\\
60489 & 665.645 & 53.42 & 0.69\\
60489 & 692.608 & 30.81 & 0.48\\
60489 & 693.650 & 31.65 &0.28\\
\\
62437 & 663.627 & 11.13 & 0.48\\
62437 & 665.677 & 14.46 & 1.34\\
62437 & 692.628 & 15.99 &0.28\\
62437 & 693.639 & 14.12 & 0.69\\
\hline
\end{tabular}
\end{flushleft}
\end{table}

\begin{table}
\hspace{0cm}{\bf Table 3 (continued).} 
\begin{flushleft}
\begin{tabular}{rcrc}
\hline
HD  & HJD     & $V_r$ & $I$ \\
    & (+2449000)&km\,s$^{-1}$&km\,s$^{-1}$\\
\hline
69997 & 663.642 & 31.67 &0.28\\
69997 & 692.641 & 32.79 &0.28\\
69997 & 692.703 & 33.18 &0.28\\
69997 & 693.661 & 32.59 &0.28\\
\\
82043 & 663.652 & 12.25 & 0.48\\
82043 & 692.649 &  1.73 & 0.69\\
82043 & 692.712 &  1.41 & 0.69\\
82043 & 693.679 &  0.14 & 0.48\\
\\
84607 & 663.660 & 13.65 & 0.48\\
84607 & 692.660 & 12.25 & 0.69\\
84607 & 692.723 & 13.50 & 0.69\\
84607 & 693.690 & 13.09 & 0.48\\
\\
86611 & 663.688 & 26.47 & 0.90\\
86611 & 692.669 & 24.42 & 0.90\\
86611 & 692.732 & 25.79 & 0.69\\
86611 & 693.699 & 25.04 & 1.34\\
\\
89025 & 692.676 &  -21.20 & 0.69\\
89025 & 693.710 &  -22.07 &0.28\\
\\
92787 & 692.689 &  4.47 & 0.90\\
92787 & 692.692 &  4.47 & 0.90\\
92787 & 693.715 &  3.82 & 0.48\\
\\
100418 & 470.406 & 1.02 & 0.69\\
100418 & 471.327 & 0.76 & 0.58 \\
100418 & 471.390 & -0.57 & 0.69\\
100418 & 472.398 & 2.28 &0.28\\
100418 & 473.336 & 0.22 & 0.69\\
100418 & 474.359 & -1.81 & 0.48\\
100418 & 506.372 & -0.28 &0.28\\
\\
103313 & 470.461 & 16.10 & 0.69\\
103313 & 471.340 & 16.75 & 0.48\\
103313 & 471.463 & 14.71 &0.28\\
103313 & 472.413 & 19.14 & 0.90\\
103313 & 473.359 & 17.59 & 0.69\\
103313 & 473.440 & 17.88 & 0.90\\
103313 & 474.378 & 16.54 & 0.69\\
103313 & 474.414 & 16.72 & 0.48\\
103313 & 505.387 & 22.17 & 1.56\\
103313 & 507.364 & 18.64 & 0.48\\
103313 & 507.397 & 15.74 & 2.01\\
\\
104827 & 470.475 & 10.21 & 0.48\\
104827 & 471.352 & 8.07& 0.90\\
104827 & 471.399 & 7.33 & 0.48\\
104827 & 472.443 & 7.70 & 0.48\\
104827 & 473.373 & 7.60 & 0.48\\
104827 & 473.482 & 8.42 & 0.48\\
104827 & 474.332 & 7.37 & 0.69\\
104827 & 474.479 & 6.93 & 0.48\\
\hline
\end{tabular}
\end{flushleft}
\end{table}

\begin{table}
\hspace{0cm}{\bf Table 3 (continued).} 
\begin{flushleft}
\begin{tabular}{rcrc}
\hline
HD  & HJD     & $V_r$ & $I$ \\
    & (+2449000)&km\,s$^{-1}$&km\,s$^{-1}$\\
\hline
104827 & 506.394 & 8.89 & 0.48\\
\\
108382 & 470.486 & -2.33 & 0.48\\
108382 & 471.364 & -1.97 & 0.48\\
108382 & 471.471 & -1.82 & 0.48\\
108382 & 472.457 & -2.86 & 0.69\\
108382 & 473.383 & -1.97 & 0.90\\
108382 & 473.430 & -2.71 & 0.69\\
108382 & 474.345 & -1.55 & 0.90\\
\\
118295 & 470.535 &   -20.00 & 0.69\\
118295 & 471.375 &   -19.63 & 0.69\\
118295 & 471.447 &   -17.52 & 1.56\\
118295 & 472.470 &   -20.59 & 0.69\\
118295 & 473.401 &   -18.46 & 0.94\\
118295 & 474.394 &   -17.76 & 0.69\\
118295 & 505.410 &  -24.23 & 0.94\\
118295 & 506.420 &   -19.84 & 1.34\\
108382 & 474.446 & -1.89 & 0.90\\
\\
122703 & 470.522 &   -17.63 & 0.69\\
122703 & 471.430 &   -13.34 & 0.90\\
122703 & 471.485 &   -21.60 & 0.69\\
122703 & 472.509 &   -19.21 & 2.01\\
122703 & 473.455 &   -15.39 & 0.69\\
122703 & 474.464 &   -17.16 & 0.94\\
\\
150453 & 471.521 & 0.82 & 0.69\\
150453 & 471.620 & 0.47 & 0.69\\
150453 & 472.586 & 0.12 &0.28\\
150453 & 473.541 & 0.64 & 0.48\\
150453 & 474.568 & -0.10 &0.28\\
150453 & 506.486 & 1.54 &0.28\\
\\
150557 & 471.510 &   -49.75 & 0.48\\
150557 & 472.492 &   -49.42 & 0.69\\
150557 & 473.496 &   -48.36 & 0.90\\
150557 & 474.494 &   -49.64 & 0.69\\
150557 & 474.540 &   -50.22 & 0.48\\
150557 & 505.532 &   -48.20 & 0.48\\
150557 & 506.412 &   -48.49 & 0.94\\
150557 & 506.467 &   -50.15 &0.28\\
150557 & 509.393 &   -49.60 & 0.48\\
\\
155646 & 470.552 & 59.76 & 0.48\\
155646 & 471.535 & 61.08 & 0.48\\
155646 & 472.542 & 62.36 & 0.48\\
155646 & 473.518 & 59.78 &0.28\\

155646 & 474.524 & 58.64 & 0.48\\
155646 & 505.545 & 60.05 & 0.48\\
155646 & 506.445 & 59.95 &0.28\\
155646 & 506.547 & 60.70 &0.28\\
\\
159561 & 470.565 & 10.69 & 0.69\\
159561 & 471.543 & 11.10 & 0.48\\
\hline
\end{tabular}
\end{flushleft}
\end{table}

\begin{table}
\hspace{0cm}{\bf Table 3 (continued).} 
\begin{flushleft}
\begin{tabular}{rcrc}
\hline
HD  & HJD     & $V_r$ & $I$ \\
    & (+2449000)&km\,s$^{-1}$&km\,s$^{-1}$\\
\hline
159561 & 472.550 & 12.23 & 0.69\\
159561 & 472.631 & 10.25 & 0.94\\
159561 & 473.527 & 12.03 & 0.90\\
159561 & 474.510 & 10.82 & 0.69\\
159561 & 505.431 & 11.46 & 0.90\\
159561 & 506.457 & 10.65 & 0.48\\
159561 & 507.608 & 9.68 & 0.69\\
159561 & 509.446 & 8.29 & 0.48\\
\\
171856 & 470.639 & -5.56 & 2.01\\
171856 & 473.620 & -5.96 & 1.34\\
171856 & 506.584 &   -21.70 & 0.48\\
171856 & 508.544 &   -19.14 & 0.69\\
\\
174866 & 470.625 &   -39.39 & 0.94\\
174866 & 472.615 &   -38.50 & 0.69\\
174866 & 473.606 &   -40.45 & 0.48\\
174866 & 505.585 &   -40.89 & 0.90\\
174866 & 506.571 &   -43.56 & 1.34\\
174866 & 509.521 &   -42.08 & 0.69\\
\\
176971 & 470.575 &   -34.62 & 0.90\\
176971 & 471.553 &   -34.28 & 0.48\\
176971 & 472.559 &   -31.63 & 0.69\\
176971 & 473.560 &   -35.13 & 0.48\\
176971 & 474.579 &   -33.63 & 0.69\\
176971 & 474.625 &   -34.88 & 1.34\\
176971 & 505.461 &   -34.60 & 0.94\\
176971 & 506.517 &   -32.51 & 0.69\\
176971 & 509.414 &   -29.47 & 1.78\\
\\
177392 & 470.608 & 10.13 & 0.48\\
177392 & 471.577 &  9.51 & 1.56\\
177392 & 472.603 & 12.38 & 0.69\\
177392 & 473.580 & 10.12 & 0.90\\
177392 & 474.593 & 6.75 & 0.90\\
177392 & 505.481 & 11.21 & 0.90\\
177392 & 506.502 & 10.31 & 2.01\\
177392 & 508.525 & 6.75 & 0.90\\
177392 & 509.459 & 8.69 &0.28\\
177392 & 509.539 & 7.37 & 0.90\\
\\
178187 & 470.600 &   -24.85 & 0.69\\
178187 & 471.563 &   -24.43 & 0.69\\
178187 & 472.568 &   -24.82 & 0.48\\
178187 & 473.570 &   -25.11 & 0.94\\
178187 & 474.602 &   -24.69 & 0.90\\
178187 & 505.440 &   -24.69 & 0.69\\
178187 & 505.518 &   -25.04 & 0.90\\
178187 & 506.531 &  -25.03 & 0.90\\
178187 & 507.585 &  -22.70 & 0.48\\
178187 & 508.506 &   -23.06 & 0.69\\
\\
186005 & 472.623 &   -40.39 & 1.34\\
186005 & 474.634 &   -39.98 & 0.90\\
\hline
\end{tabular}
\end{flushleft}
\end{table}

\begin{table}
\hspace{0cm}{\bf Table 3 (continued).} 
\begin{flushleft}
\begin{tabular}{rcrc}
\hline
HD  & HJD     & $V_r$ & $I$ \\
    & (+2449000)&km\,s$^{-1}$&km\,s$^{-1}$\\
\hline
186005 & 506.593 &   -39.50 & 0.90\\
186005 & 509.593 &   -42.63 & 0.48\\
\\
187764 & 471.598 & -8.02 & 0.69\\
187764 & 473.589 & -5.96 & 0.48\\
187764 & 474.551 & -8.39 & 0.90\\
187764 & 474.612 & -4.71 & 0.90\\
187764 & 505.560 & -6.14 & 0.69\\
187764 & 506.554 & -3.10 & 0.69\\
187764 & 509.477 & -4.64 & 0.69\\
\\
190172 & 471.628 & 2.82 & 0.48\\
190172 & 473.628 & 6.00 &0.28\\
190172 & 506.602 & 3.11 & 0.48\\
\\
203842 & 663.228 &  -25.94 & 0.69\\
203842 & 665.246 &  -27.74 & 0.48\\
203842 & 690.313 &  -26.43 & 0.69\\
203842 & 692.296 &  -24.73 & 0.90\\
203842 & 693.243 &  -23.22 & 0.48\\
\\
204577 & 663.245 &  -10.45 & 0.48\\
204577 & 693.273 & -8.69 & 0.69\\
\\
205852 & 663.262 &  -31.90 & 0.90\\
205852 & 692.307 &  -31.83 & 0.69\\
205852 & 693.308 &  -32.48 & 0.69\\
\\
209166 & 663.275 &  5.22 &0.28\\
209166 & 692.314 &  6.36 &0.28\\
209166 & 693.322 &  6.82 & 0.48\\
\\
210516 & 663.295 & 10.46 & 0.69\\
210516 & 692.332 & 11.38 & 0.94\\
210516 & 693.339 & 12.16 & 0.94\\
\\
216701 & 663.314 & 12.70 & 0.48\\
216701 & 692.352 & 15.26 & 0.69\\
216701 & 693.355 & 11.36 & 0.48\\
\\
217131 & 663.334 &  -12.12 &0.28\\
217131 & 693.374 &  -11.35 & 0.48\\
\\
219891 & 663.352 & -3.04 & 0.90\\
219891 & 689.338 & -3.01 & 2.01\\
219891 & 693.396 & -0.48 & 0.94\\
\\
224995 & 663.383 &  8.13 & 0.48\\
224995 & 689.302 &  7.14 & 0.90\\
224995 & 693.431 &  7.48 & 0.69\\
\hline
\vspace{15cm}
\end{tabular}
\end{flushleft}
\end{table}

%\end{document}

\noindent
{\bf Table 6.} Discussion about the duplicity of all photometric metallic F
giant stars of Hauck  (1986)

\vspace{2mm}
\begin{minipage}[t]{2cm}HD 1324
\end{minipage}
\begin{minipage}[t]{8.8cm}Single star (Evans et al. 1964).
\end{minipage}

\vspace{3mm}
\begin{minipage}[t]{2cm}HD 2724
\end{minipage}
\begin{minipage}[t]{8.8cm}$\delta$ Scuti (Rodriguez et al. 1994). The $V_{r}$
measurements of Nordström \& Andersen (1985) show no variation.
\end{minipage}

\vspace{3mm}
\begin{minipage}[t]{2cm}HD 4849
\end{minipage}
\begin{minipage}[t]{8.8cm}$\delta$ Scuti (Rodriguez et al. 1994). The $V_{r}$
measurements of Nordström \& Andersen (1985) show no variation.
\end{minipage}

\vspace{3mm}
\begin{minipage}[t]{2cm}HD 4919
\end{minipage}
\begin{minipage}[t]{8.8cm} $\delta$ Scuti (Rodriguez et al. 1994). As we only
have an old $V_{r}$ from the catalogue of Campbell (1913), we cannot conclude
about the duplicity.
\end{minipage}

\vspace{3mm}
\begin{minipage}[t]{2cm}HD 10845
\end{minipage}
\begin{minipage}[t]{8.8cm} $\delta$ Scuti (Rodriguez et al. 1994). Our
measurements show no variation.
\end{minipage}

\vspace{3mm}
\begin{minipage}[t]{2cm}HD 12311
\end{minipage}
\begin{minipage}[t]{8.8cm} From his 15 exposures, Campbell (1928) indicated
that this star is probably variable. At the time, he had only a few broad lines. 
Catchpole et al. (1982) observed no variation from 4 measurements. They obtained
a mean radial velocity of 15.2 $\pm$3.1 km\,s$^{-1}$. 
\end{minipage}

\vspace{3mm}
\begin{minipage}[t]{2cm}HD 13174
\end{minipage}
\begin{minipage}[t]{8.8cm} Campbell (1928) obtained values covering a large
interval  [-10,+20 km\,s$^{-1}$]. This dispersion is probably due essentially
to a fast rotation ($v\sin i$ =154), nevertheless we cannot exclude a close
companion. Adams et al (1929) and Abt (1969) gave coherent velocities of about
-6 km\,s$^{-1}$. 
 \end{minipage}

\vspace{3mm}
\begin{minipage}[t]{2cm}HD 15233
\end{minipage}
\begin{minipage}[t]{8.8cm} The comparison between values obtained by Campbell
(1913,1928) and by Buscombe \& Morris (1958) shows that this star must be a
binary with a long period. Indeed, these authors gave very different radial
velocities: one found positive velocities, the others negative velocities,
with a very good internal coherence if we consider the high rotation
($v\sin i$ = 106 km\,s$^{-1}$). We find a value of $P(\chi^2)$ of 0.08 for
the measurements of Buscombe \& Morris (1958) showing, according to this
criterion, that no variation is detected on a scale of 700 days. 

\end{minipage}

\vspace{3mm}
\begin{minipage}[t]{2cm}HD 17584
\end{minipage}
\begin{minipage}[t]{8.8cm} Our measurements show no variation. We obtain a
velocity of 18 km\,s$^{-1}$ and Campbell (1913) 14 km\,s$^{-1}$.
\end{minipage}

\vspace{3mm}
\begin{minipage}[t]{2cm}HD 17918
\end{minipage}
\begin{minipage}[t]{8.8cm} Our measurements show no variation as is the case of
the measurements of Shajn \& Albitzky (1932). 
\end{minipage}

\vspace{3mm}
\begin{minipage}[t]{2cm}HD 30020
\end{minipage}
\begin{minipage}[t]{8.8cm} The measurements made at OHP are compatible with the
preceding ones of Adams et al (1929) and Abt (1969). The available values are
distributed between 34.40 km\,s$^{-1}$ and 41.50 km\,s$^{-1}$ with a
characteristic time  variation of about 0.7 km\,s$^{-1}$/day taking into account
our measurements of December 1994. The variability is certainly due to an orbit
because the photometric variations are less than 1 mmag in the filter B and V
of Johnson (Nelson \& Kreidl 1993).
\end{minipage}

\vspace{6mm}

\noindent
{\bf Table 6.} (Continued)

\vspace{3mm}
\begin{minipage}[t]{2cm}HD 33276
\end{minipage}
\begin{minipage}[t]{8.8cm}The measurements of Campbell (1928) and Abt (1965)
show no variation. 
\end{minipage}

\vspace{3mm}
\begin{minipage}[t]{2cm}HD 34045
\end{minipage}
\begin{minipage}[t]{8.8cm}If we put together our measurements and those of  
Nordström \& Andersen (1985), the velocity varies between 26.32 and
34.70 km\,s$^{-1}$. The characteristic time of variation is
0.95 km\,s$^{-1}$/day for our measurements (3 days between two measurments) and 
0.91 km\,s$^{-1}$/day for Nordström \& Andersen (1985) (5 days between two 
measurements). As this star is not known as a $\delta$ Scuti, we think it is a 
tight binary. 
\end{minipage}

\vspace{3mm}
\begin{minipage}[t]{2cm}HD 40455
\end{minipage}
\begin{minipage}[t]{8.8cm}For this star, we have only 3 measurements of 
Nordström \& Andersen (1985) on a scale of 300 days. The $P(\chi^2)$ is 0.74 
and so this star is probably not a tight binary.
\end{minipage}

\vspace{3mm}
\begin{minipage}[t]{2cm}HD 43905
\end{minipage}
\begin{minipage}[t]{8.8cm}SB1 with a period of 6.5 days (Mayor \& Mazeh 1987).
\end{minipage}

\vspace{3mm}
\begin{minipage}[t]{2cm}
HD 59881
\end{minipage}
\begin{minipage}[t]{8.8cm}
The $P(\chi^2)$ is 0.06 and 0.13 for the measurements of  Frost et al (1929)
and Penfold (1983) respectively. So this star is probably not a binary with a
small period.
\end{minipage}

\vspace{3mm}
\begin{minipage}[t]{2cm}
HD 61064
\end{minipage}
\begin{minipage}[t]{8.8cm}
If we put together the measurements of Campbell (1913,1928) and those of
Shajn \& Albitzky (1932), the mean value of the radial velocity is
45.72 km\,s$^{-1}$ with a dispersion of 1.72 km\,s$^{-1}$, so we observe
no tangible variation.
\end{minipage}

\vspace{3mm}
\begin{minipage}[t]{2cm}
HD 61110
\end{minipage}
\begin{minipage}[t]{8.8cm}
The measurements taken from Campbell (1913,1928), Harper (1937),
Buscombe \& Morris (1958), Abt (1969) give a mean radial velocity of
6.5 km\,s$^{-1}$ and a dispersion of 3.5 km\,s$^{-1}$ for 13 data. The 
dispersion is not so large if we consider the quite large rotational velocity 
(90 km\,s$^{-1}$). So this star is probably not a tight binary.
\end{minipage}

\vspace{3mm}
\begin{minipage}[t]{2cm}
HD 69997
\end{minipage}
\begin{minipage}[t]{8.8cm}
$\delta$ Scuti (Rodriguez et al. 1994). Our measurements allow to detect 
intrinsic variations but no orbital motion.
\end{minipage}

\vspace{3mm}
\begin{minipage}[t]{2cm}
HD 79940
\end{minipage}
\begin{minipage}[t]{8.8cm}
Single star (Evans et al. 1961). This star is noted SB in the BSC. This
probably comes from Abt \& Biggs (1972). Indeed, in addition to the authors
mentioned above who find a radial velocity of $2.2 \pm 1.5$ km\,s$^{-1}$, they
list Campbell (1928, p. 143) who found a radial velocity of 12 km\,s$^{-1}$
based on 3 measurements. The large rotational velocity probably explains the
discrepancy.
\end{minipage}

\vspace{20mm}

\noindent
{\bf Table 6.} (Continued)

\vspace{3mm}
\begin{minipage}[t]{2cm}
HD 84607
\end{minipage}
\begin{minipage}[t]{8.8cm}
We detect no variation of this star at OHP. We obtain a mean radial velocity
of 13 km\,s$^{-1}$ which is compatible with the value of 16.3 $\pm$ 2 given
by Shajn \& Albitzky (1932). Wilson \& Joy (1950) and Abt (1969) both found
important fluctuations of velocity, which are spurious, probably due to the
rotation. In both cases, the mean value is 13.4 km\,s$^{-1}$. The SB notation
in the BSC seems to be a premature decision.
\end{minipage}

\vspace{3mm}
\begin{minipage}[t]{2cm}
HD 85040
\end{minipage}
\begin{minipage}[t]{8.8cm}
$\delta$ Scuti (Rodriguez et al. 1994). SB2 with a period of 4.14 days
(Rosvick \& Scarfe 1991)
\end{minipage}

\vspace{3mm}
\begin{minipage}[t]{2cm}
HD 89025
\end{minipage}
\begin{minipage}[t]{8.8cm}
The data since the beginning of the century show a radial velocity variation
about 50 km\,s$^{-1}$, but no short-term variation is detected, so it is
probably not a tight binary as is suggested in the BSC. Indeed, apart from
the values of Henroteau (1923) who obtained a mean velocity of about
-54 km\,s$^{-1}$, the others have radial velocities between -30 and
0 km\,s$^{-1}$ with a mean of -17 km\,s$^{-1}$. This scatter is probably due
both to the quite large $v\sin i$  (81 km\,s$^{-1}$) and the low dispersion of
the spectra ( $\approx$ 30\AA/mm). The computation of the $P(\chi^2)$ gives
0.88 and 0.53 for the measurements of Jones \& Haslam (1969) and
Wooley et al (1971) respectively, and 0 for the old measurements of Abt (1969).
Our radial velocities are compatible with those of Adams et al (1929) who
obtained -19.3 $\pm$ 2.9.
\end{minipage}

\vspace{3mm}
\begin{minipage}[t]{2cm}
HD 89254
\end{minipage}
\begin{minipage}[t]{8.8cm}
The measurements given by Adams et al. (1929), Buscombes \& Morris (1958),
Abt (1969) and Evans et al. (1961) suggest a constant radial velocity 
($\overline{V_{r}}=15.8\pm 4.4$). This dispersion is compatible with a 
rotational velocity of 76 km\,s$^{-1}$.
\end{minipage}

\vspace{3mm}
\begin{minipage}[t]{2cm}
HD 90277
\end{minipage}
\begin{minipage}[t]{8.8cm}
This star shows no variation. The mean velocity is 13.6 km\,s$^{-1}$ with a
dispersion of 2.7 km\,s$^{-1}$ from the 36 measurements taken from 9 different
authors.
\end{minipage}

\vspace{3mm}
\begin{minipage}[t]{2cm}
HD 90454
\end{minipage}
\begin{minipage}[t]{8.8cm}
From the measurements of Cannon (1920) and those of
Nordström \& Andersen (1985), the $P(\chi^2)$ is 0.31 and 0.03 respectively and
so it is probably not a binary. 
\end{minipage}

\vspace{3mm}
\begin{minipage}[t]{2cm}
HD 105841
\end{minipage}
\begin{minipage}[t]{8.8cm}
Single star (Evans 1966).
\end{minipage}

\vspace{3mm}
\begin{minipage}[t]{2cm}
HD 108722
\end{minipage}
\begin{minipage}[t]{8.8cm}
SB1 with a period of 17.954 days (Abt \& Levy 1976)
\end{minipage}

\vspace{3mm}
\begin{minipage}[t]{2cm}
HD 110318
\end{minipage}
\begin{minipage}[t]{8.8cm}
SB2 with a period of 44.4 days (Sanford \& Karr 1942)
\end{minipage}

\vspace{3mm}
\begin{minipage}[t]{2cm}
HD 115604
\end{minipage}
\begin{minipage}[t]{8.8cm}
$\delta$ Scuti (Rodriguez et al. 1994). On a time span of about 130 days,
M.A. Smith (1982) observes a variation of radial velocity of 6.2 km\,s$^{-1}$.
Such a variation is probably due to pulsation only.
\end{minipage}

\vspace{3mm}
\begin{minipage}[t]{2cm}
HD 125150
\end{minipage}
\begin{minipage}[t]{8.8cm}
Single star (Evans et al 1964).
\end{minipage}

\vspace{20mm}

\noindent
{\bf Table 6.} (Continued)

\vspace{3mm}
\begin{minipage}[t]{2cm}
HD 126251
\end{minipage}
\begin{minipage}[t]{8.8cm}
Single star (Wilson \& Joy 1950).
\end{minipage}

\vspace{3mm}
\begin{minipage}[t]{2cm}
HD 147547
\end{minipage}
\begin{minipage}[t]{8.8cm}
Radial velocities are distributed between -60 and -30 km\,s$^{-1}$ with a
mean of about -44 km\,s$^{-1}$. The large dispersion is probably due to the
fast rotation of 145 km\,s$^{-1}$, nevertheless we cannot exclude a
contribution of a companion.
\end{minipage}

\vspace{3mm}
\begin{minipage}[t]{2cm}
HD 157919
\end{minipage}
\begin{minipage}[t]{8.8cm}
It shows no variation. The mean radial velocity is 37.2$\pm$2.4 km\,s$^{-1}$
for 14 measurements taken from Campbell (1913), Campbell (1928),
Shajn \& Albitzky (1932), Evans et al. (1957)
\end{minipage}

\vspace{3mm}
\begin{minipage}[t]{2cm}
HD 177392
\end{minipage}
\begin{minipage}[t]{8.8cm}
$\delta$ Scuti (Rodriguez et al. 1994). Our measurements allow to detect
intrinsic variations but no orbital motion.
\end{minipage}

\vspace{3mm}
\begin{minipage}[t]{2cm}HD 177482
\end{minipage}
\begin{minipage}[t]{8.8cm}
$\delta$ Scuti (Rodriguez et al. 1994). Campbell (1928) obtains values
covering a large interval [+3,+26 km\,s$^{-1}$] with the remark ``poor lines''
and Neubauer (1930) gives a value of 6.5 km\,s$^{-1}$. With these old values,
it is difficult to say anything about the duplicity.
\end{minipage}

\vspace{3mm}
\begin{minipage}[t]{2cm}
HD 181333
\end{minipage}
\begin{minipage}[t]{8.8cm}
$\delta$ Scuti (Rodriguez et al. 1994). On a time span of about 38 days,
M.A. Smith (1982) observes a variation of radial velocity of 4.7 km\,s$^{-1}$.
Such a variation is probably due to pulsation only.
\end{minipage}

\vspace{3mm}
\begin{minipage}[t]{2cm}
HD 196524
\end{minipage}
\begin{minipage}[t]{8.8cm}
Binary star with a probable period of 26.65 years (Abt \& Levy 1976). So it
will not be integrated in the sample of close binaries. 
\end{minipage}

\vspace{3mm}
\begin{minipage}[t]{2cm}
HD 205852
\end{minipage}
\begin{minipage}[t]{8.8cm}
From the $P(\chi^2)$ value, the measurements of Frost et al (1929),
Jones \& Haslam (1969) and this paper show no variations.
\end{minipage}

\vspace{3mm}
\begin{minipage}[t]{2cm}
HD 208741
\end{minipage}
\begin{minipage}[t]{8.8cm}
Single star (Buscombe \& Morris 1958 and Evans et al 1964).
\end{minipage}

\vspace{3mm}
\begin{minipage}[t]{2cm}
HD 214441
\end{minipage}
\begin{minipage}[t]{8.8cm}
$\delta$ Scuti with a variation of 0.05 mag in visual magnitude
(Rodriguez et al. 1994). Nordström \& Andersen (1985) find a $\Delta V_r$ of
about 10 $kms^{-1}$ and  we then obtain a ratio of 200 $kms^{-1}mag^{-1}$
which is large in comparison with the value of 92 $kms^{-1}mag^{-1}$ given
by Breger (1979). We cannot therefore exclude a companion for this star.
\end{minipage}


\begin{thebibliography}{}
\bibitem{} Abt H. A., 1965, ApJS 11, 429
\bibitem{} Abt H. A., 1969, ApJS 19, 387
\bibitem{} Abt H. A., Biggs E. S., 1972, Bibliography of Stellar Radial
Velocities, Latham Process Corp., New York
\bibitem{} Abt H. A., Levy S. G., 1976, ApJS 30, 273
\bibitem{} Abt H. A., Levy S. G., 1985, ApJS 59, 229
\bibitem{} Abt H. A., Morrell N. I., 1995, ApJS 99, 135
\bibitem{} Abt H. A., Moyd K. I., 1973, ApJ 182, 809
\bibitem{} Adams W. S., 1923, ApJ 57, 149
\bibitem{} Adams W. S., Joy A. H., Sanford R. F., Strömberg G., 1929,
ApJ 70, 207
\bibitem{} Alecian G., 1996, A\&A 310, 878
\bibitem{} Baranne A., Mayor M., Poncet J.-L., 1979, Vistas Asron. 23, 279
\bibitem{} Bernacca P.L., Perinotto M., 1971, Contr. Oss. astrofis. Univ.
Padova 249
\bibitem{} Berthet S., 1990, A\&A 227, 156
\bibitem{} Berthet S., 1991, A\&A 251, 171
\bibitem{} Berthet S., 1992, A\&A 253, 451
\bibitem{} Boyarchuk A.A., Kopylov I.M., 1964, Publ. Obs. Crimee, 31, 44
\bibitem{} Breger M., 1979, PASP 91, 5
\bibitem{} Buscombe W., Morris P. M., 1958, MNRAS 118, 609
\bibitem{} Campbell W. W., 1911, Lick Obs. Bull. 6, 140
\bibitem{} Campbell W. W., 1912, Lick Obs. Bull. 7, 19
\bibitem{} Campbell W. W., 1913, Lick Obs. Bull. 7, 113
\bibitem{} Campbell W. W., 1928, Pub. of the Lick Obs. 16, 1
\bibitem{} Cannon J. B.,1920, Pub. of the Dominion Obs. 4, 253
\bibitem{} Catchpole R. M., Evans D. S., Jones D. H. P., King D. L.,
Wallis R. E., 1982, Royal Greenwich Obs. Bull. 188, 5
\bibitem{} Charbonneau P., Michaud G., 1991, ApJ 370, 693
\bibitem{} Cochran W.D., Hatzes A.P., Hancock T.J., 1991, ApJ, 380, 35
\bibitem{} Cowley A.P., 1976, PASP, 88, 95
\bibitem{} Cowley A.P., Cowley C.R., Jaschek M., Jaschek C., 1969, AJ, 74, 375
\bibitem{} Curchod A., Hauck B., 1979, A\&AS 38, 449
\bibitem{} Duquennoy A., Mayor, M. 1991, A\&A 248, 485
\bibitem{} Evans D. E., 1966, Royal Greenwich Obs. Bull. 110, 185
\bibitem{} Evans D. E., Laing J. D., Menzies A., Stoy R. H., 1964,
Royal Greenwich Obs. Bull. 85, 207
\bibitem{} Evans D. E., Menzies A., Stoy R. H., 1957, MNRAS 117, 534
\bibitem{} Evans D. E., Menzies A., Stoy R. H.,Wayman P. A., 1961,
Royal Greenwich Obs. Bull. 48, 389
\bibitem{} Frost E. B., Barrett S. B., Struve O., 1929, Pub. of the
Yerkes Obs. 7, 1
\bibitem{} Gillet D., Burnage R., Kohler D. et al, 1994, A\&AS 108, 181
\bibitem{} Harper W. E., 1920, Pub. of the Dominion Obs. 4 , 331
\bibitem{} Harper W. E., 1934, Pub. of the Dominion Astrophys. Obs. 6, 151
\bibitem{} Harper W. E., 1937, Pub. of the Dominion Astrophys. Obs. 7 , 1
\bibitem{} Hauck B., 1973, problems of calibration of Absolute Magnitudes
and Temperature of Stars, IAU Symp. 54, 117
\bibitem{} Hauck B., 1986, A\&A 155, 371
\bibitem{} Hauck B., Curchod A., 1980, A\&A 92, 289
\bibitem{} Hauck B., Jaschek C., Jaschek M, Andrillat Y., 1991, A\&A 252, 260
\bibitem{} Henroteau F., 1923, Pub. of the Dominion Obs. 8, 59
\bibitem{} Hoffleit D., Jaschek C., 1982, The Bright Star Catalogue, 4th
revised edition, Yale University Observatory, New Haven
\bibitem{} Houk N., Cowley A.P., 1975, Michigan Catalogue of Two-Dimensional
Spectral Types for the HD Stars, Univ. of Michigan, Ann Arbor, Vol.1
\bibitem{} Houk N., 1978, Michigan Catalogue of Two-Dimensional Spectral Types
for the HD Stars, Univ. of Michigan, Ann Arbor, Vol.2
\bibitem{} Houk N., 1982, Michigan Catalogue of Two-Dimensional Spectral Types 
for the HD Stars, Univ. of Michigan, Ann Arbor, Vol.3
\bibitem{} Jaschek M., 1978, A Catalogue of Selected MK Types, Inf. Bull. CDS
15, 121
\bibitem{} Jones H. P., Haslam C. M., 1969 Royal Greenwich Obs. Bull., 155
\bibitem{} Jordan F. C., 1912, Pub. of the Allegheny Obs. 2, 121
\bibitem{} Künzli M., North P., Kurucz R. L., Nicolet B., 1996, A\&AS 122, 51
\bibitem{} Kurtz D. W., 1976, ApJS 32, 651
\bibitem{} Latham D.W., Mazeh T., Stefanik R.P., Mayor M., Burki G.,
1989, Nature, 339, 38
\bibitem{} Mayor M., Maurice E., 1985, in: Stellar radial velocities, IAU Coll.
No. 88, eds. A.G. Davis Philip and David W. Latham, Davis Press, p. 299
\bibitem{} Mayor M., Mazeh T., 1987, A\&A 171, 157
\bibitem{} Mayor M., Udry, S., 1996, private communication
\bibitem{} Mazeh T., Goldberg D., Duquennoy A., Mayor M., 1992, ApJ 401, 265
\bibitem{} Michaud G., Charland Y., Vauclair S., Vauclair G., 1976, ApJ 210, 447
\bibitem{} Michaud G., Tarasick D., Charland Y., Pelletier C., 1983, ApJ 269,
239
\bibitem{} Nelson J.N., Kreidl T.J., 1993, AJ, 105, 1903
\bibitem{} Neubauer F. J., 1930, Lick Obs. Bull. 15, 46
\bibitem{} Nordström B., Andersen J., 1985, A\&AS 61 , 53
\bibitem{} North P., 1994, in: The 25th workshop and meeting of European
working group on CP stars, eds. I. Jankovics and I.J. Vincze, Gothard
Astrophysical Observatory of E\"otv\"os University, Szombathely, Hungary, p. 3
\bibitem{} 0'Brien G. T. et al, 1986, ApJS 62 , 899
\bibitem{} Penfold J.E., 1971, PASP 83, 497
\bibitem{} Plaskett W.E., Harper W. E., Young R. K., Plaskett H. H., 1921,
Pub. of the Dominion Astrophys. Obs., 2, 1. Spencer Jones, H., 1928,
CAPE 10-8, 95
\bibitem{} Rodriguez E., Lopez de Coca P., Rolland A., Garrido R., Costa V.,
1994, A\&AS 106, 21
\bibitem{} Rosvick J. M., Scarfe C. D., 1991, PASP 103, 628
\bibitem{} Sanford R.F., Karr E., 1942, ApJ 96, 214
\bibitem{} Shajn G., Albitzky V., 1932, MNRAS 92, 771
\bibitem{} Smith M.A., 1982, ApJ, 254, 242
\bibitem{} Tassoul J.-L., Tassoul M., 1992, ApJ, 395, 259
\bibitem{} Uesugi A., Fukuda I., 1978, Preliminary Catalogue of Rotational
velocities, CDS
\bibitem{} Uesugi A., Fukuda I., 1982, Revised Catalogue of Rotational
Velocities, Uni. of Kyoto
\bibitem{} Vauclair S., Vauclair G., 1982, Ann. Rev. Astron. Astrophys. 20, 37
\bibitem{} Wilson R. E., Joy A. H., 1950, ApJ 111, 221
\bibitem{} Wooley R., Penston M. J., Harding G. A., Martin W. L., Sinclair C.
M., Aslan S., Savage A., Aly K., Assad A. S., 1971, Royal Greenwich Observatory
Annals 14, 1
\bibitem{} Zahn J. P., 1977, A\&A 57, 383
\end{thebibliography}
\end{document}